\documentclass{article} 
\usepackage{nips13submit_e,times}
\usepackage{hyperref}
\usepackage{url}

\usepackage{comment}
\usepackage{amsmath}
\usepackage{graphicx}
\usepackage{natbib}
\usepackage[noabbrev,capitalise]{cleveref}
\usepackage{lipsum}
\usepackage{amssymb}
\usepackage{multicol}
\usepackage{bm}
\usepackage{bbm}
\usepackage{upgreek}
\usepackage{bigints}
\usepackage[justification=centering]{caption}
\usepackage{wrapfig}
\usepackage{tikz}
\usepackage{float}
\usepackage{subcaption}
\usepackage{mathtools}
\usepackage{amsfonts}

\usepackage{algorithm}
\usepackage{algpseudocode}
\usepackage{amsthm}
\usepackage{booktabs}
\usepackage{listings}
\usepackage{tcolorbox}
\tcbuselibrary{listingsutf8, breakable, skins} 
\newtcblisting{rcode}{
  listing only,
  colback=gray!5,
  colframe=gray!50,
  listing options={language=R, 
  keywords={},
  keywordstyle=,
  breaklines=true,
  basicstyle=\ttfamily\footnotesize},
  left=5pt,
  right=5pt,
  top=5pt,
  bottom=5pt,
  sharp corners,
  breakable
}
\usepackage{fancyvrb}

\crefname{prop}{Proposition}{Propositions}

\usepackage{xcolor}
\definecolor{forestgreen}{rgb}{0.13, 0.55, 0.13}

\usepackage{soul}

\usepackage{microtype}
\newcommand{\tf}[1]{\textls[60]{\textup{#1}}}
\newcommand{\emu}[1]{\textsc{#1}}

\usepackage{setspace}
\title{All Emulators are Wrong, Many are Useful, and Some are More Useful Than Others: A Reproducible Comparison of Computer Model Surrogates}

\author{Kellin N. Rumsey$^1$ \\
Statistical Sciences\\
Los Alamos National Laboratory \\
Los Alamos, NM 87545 \\
\And
Graham C. Gibson \\
Statistical Sciences \\
Los Alamos National Laboratory \\
Los Alamos, NM 87545 \\
\And
Devin Francom \\
Statistical Sciences \\
Los Alamos National Laboratory \\
Los Alamos, NM 87545 \\
\And
Reid Morris \\
Statistical Sciences \\
Los Alamos National Laboratory \\
Northeastern University
}

\usepackage[normalem]{ulem}
\nipsfinalcopy 

\begin{document}

\maketitle

\normalsize

\onehalfspacing


\begin{abstract}
Accurate and efficient surrogate modeling is essential for modern computational science, and there are a staggering number of emulation methods to choose from. With new methods being developed all the time, comparing the relative strengths and weaknesses of different methods remains a challenge due to inconsistent benchmarking practices and (sometimes) limited reproducibility and transparency. In this work, we present a large-scale, fully reproducible comparison of $32$ distinct emulators across $60$ canonical test functions and $40$ real emulation datasets. To facilitate rigorous, apples-to-apples comparisons, we introduce the R package \texttt{duqling}, which streamlines reproducible simulation studies using a consistent, simple syntax, and automatic internal scaling of inputs. This framework allows researchers to compare emulators in a unified environment and makes it possible to replicate or extend previous studies with minimal effort, even across different publications. Our results provide detailed empirical insight into the strengths and weaknesses of state-of-the-art emulators and offer guidance for both method developers and practitioners selecting a surrogate for new data. We discuss best practices for emulator comparison and highlight how \texttt{duqling} can accelerate research in emulator design and application.
\end{abstract}

\doublespacing

\newpage

\standardsize

\section{Introduction}

A {\it computer model emulator} (also called a {\it surrogate} or {\it meta-model}) is a fast-to-evaluate statistical proxy for an expensive (and/or proprietary) computer model.  Emulators are essential tools in uncertainty quantification (UQ), enabling analyses that would otherwise be computationally prohibitive or infeasible.

According to \cite{lu2024survey}, there are roughly ten thousand new emulation-related papers released each year, with more than two hundred thousand papers written on the subject since the year $2000$.  We have recently witnessed the emergence of countless surveys, review papers, and even textbooks (\cite{razavi2012review, sudret2017surrogate, alizadeh2020managing, liu2020gaussian, wang2022recent, gramacy2020surrogates, sauer2023non} to name a few). When it comes to choosing an emulation method, there are dozens of possible methods, each with its own extensions and variants. Hence, it is difficult to know for certain which needle in the emulation-haystack one ought to select for a particular task. Developers of new surrogate models are expected to compare their methods against existing approaches, yet the sheer number of possible comparisons often undermines the interpretability and trustworthiness of these results. Practitioners in various communities are seldom aware of the pros and cons (or indeed, the existence) of techniques used by neighboring scientific communities.  \cite{mcclarren2011physics, salter2016comparison, laloy2019emulation, heaton2019case, hutchings2023comparing, chipman2012bart, rumsey2023localized} and \cite{collins2024bayesian} represent a few (among many) recent examples of emulation papers that include a large-scale ``bake-off" style comparison. Similar benchmarking efforts have also appeared in related areas, including hyperparameter optimization, supervised learning, and classification \citep{cowen2022hebo, fernandez2014we, kigerl2022great, marrs2025multiclass}. Not surprisingly, different papers come to different conclusions depending on the context.

In this work, we focus on the basic scenario where computer models have a vector of inputs and scalar response. Our primary goals are:
\begin{itemize}
    \item[1)] We present a framework for reproducible and transparent simulation studies, which allows for a direct comparison of emulation methods, even across papers. 
    \item[2)] We perform a massive comparison of $32$ popular methods using a diverse suite of $60$ test functions and $40$ datasets. 
    \item[3)] We offer guidance and tools for helping practitioners make informed choices about which emulator(s) they should consider for a particular UQ task. 
    \item[4)] We offer guidance and tools to help emulator methodology developers understand strengths and weaknesses (and opportunities for improvement) of various emulators.
\end{itemize}

The remainder of this paper is organized as follows. In \cref{sec:contestants}, we survey some popular approaches for emulation and discuss the $32$ methods that are the focus of the present work. In \cref{sec:duqling}, we present a framework for reproducible emulator comparisons, facilitated by the \texttt{duqling} R package. Results of this comparison are given in \cref{sec:results} with further analysis and discussion given in \cref{sec:discuss}.

\section{Review of Emulation Approaches: Meet the Contestants}
\label{sec:contestants}

\begin{table}[!htbp]
\centering
\begin{tabular}{llll}
\hline \\[-1.8ex]
{\bf Emulator} & {\bf Full Name} & {\bf Reference} & {\bf R Package} \\ \\[-1.8ex]
\hline
\\[-1.8ex]
\multicolumn{4}{l}{\textbf{Gaussian Processes and Their Variants}} \\
\emu{gp}     & Gaussian Process              & {\small\cite{rasmussen2006gaussian}}         & \texttt{hetGP} \\
\emu{rgasp}  & Robust Gaussian Process       & {\small\cite{gu2018robustgasp}}                & \texttt{RobustGaSP} \\
\emu{gpytorch}  & Gaussian Process (Python)       & {\small\cite{gardner2018gpytorch}}                & \texttt{gpytorch}$^\star$ \\
\emu{deepgp} & Deep Gaussian Process         & {\small\cite{sauer2023active}}             & \texttt{deepgp} \\
\emu{treegp} & Treed Gaussian Process        & {\small\cite{braun2024gptreeo}}           & \texttt{GPTreeO} \\
\emu{hetgp}  & Heteroskedastic GP            & {\small\cite{binois2018practical}}            & \texttt{hetGP} \\[1.2ex]

\multicolumn{4}{l}{\textbf{Approximate Gaussian Processes}} \\
\emu{mpgp}   & Matching Pursuit GP           & {\small\cite{keerthi2005matching}}           & \begin{tabular}{@{}l@{}}
\texttt{knrumsey} \\
\ \ \texttt{/spareGParts}
\end{tabular} \\
\emu{lagp}   & Local Approximate GP          & {\small\cite{gramacy2015local}}           & \texttt{laGP} \\
\emu{alcgp}  & Active Learning Cohn GP       & {\small\cite{gramacy2015local}}           & \texttt{laGP} \\
\emu{bcmgp}  & Bayesian Committee Machine    & {\small\cite{deisenroth2015distributed}}        & \begin{tabular}{@{}l@{}}
\texttt{knrumsey} \\
\ \ \texttt{/spareGParts}
\end{tabular} \\
\emu{rffgp}  & Random Fourier Features GP    & {\small\cite{rahimi2007random}}            & \texttt{gplite} \\
\emu{fitcgp} & Inducing Point GP (FITC)      & {\small\cite{snelson2005sparse}}           & \texttt{gplite} \\
\emu{svigp} & Sparse Variational GP       & {\small\cite{hensman2013gaussian}}           & \texttt{gpytorch}$^\star$ \\
\emu{svecgp} & Scaled Vecchia GP             & {\small\cite{katzfuss2022scaled}}          & \begin{tabular}{@{}l@{}}
\texttt{knrumsey} \\
\ \ \texttt{/spareGParts}
\end{tabular} \\

\multicolumn{4}{l}{\textbf{Basis Function Regression}} \\
\emu{blm}    & Bayesian Linear Model         & {\small\cite{gelman2013bayesian}}            & \texttt{stats} \\
\emu{bart}   & Bayes. Additive Regression Trees & {\small\cite{chipman2012bart}}    & \texttt{BART} \\
\emu{bass}   & Bayes. Adaptive Spline Surfaces  & {\small\cite{francom2018sensitivity}}     & \texttt{BASS} \\
\emu{tbass}  & T-distributed BASS            & {\small\cite{rumsey2024generalized}}            & \texttt{GBASS} \\
\emu{qbass}  & Quantile (Median) BASS        & {\small\cite{rumsey2024generalized}}            & \texttt{GBASS} \\
\emu{bppr}   & Bayes. Projection Pursuit Reg. & {\small\cite{collins2024bayesian}} & \begin{tabular}{@{}l@{}}
\texttt{gqcollins} \\
\ \ \texttt{/BayesPPR}
\end{tabular} \\
\emu{spce}   & Sparse Bayes. Polynomial Chaos     & {\small\cite{shao2017bayesian}}       & \texttt{khaos} \\
\emu{apce}   & Adaptive Bayes. Polynomial Chaos   & {\small\cite{rumsey2026bayesian}}     & \texttt{khaos} \\[1.2ex]

\multicolumn{4}{l}{\textbf{Machine Learning and Miscellaneous}} \\
\emu{bnn}    & Bayesian Neural Network       & {\small\cite{neal2012bayesian}}              & \texttt{bnns} \\
\emu{gbc} & Generative Bayesian Computation & {\small\cite{polson2026generative}} &
\begin{tabular}{@{}l@{}}
\texttt{VadimSokolov} \\
\ \ \texttt{/gbc-surrogate}$^\star$
\end{tabular} \\
\emu{rvm}    & Relevance Vector Machine      & {\small\cite{tipping1999relevance}}           &
\begin{tabular}{@{}l@{}}
\texttt{knrumsey} \\
\ \ \texttt{/spareGParts}
\end{tabular} \\
\emu{bootrf} & Random Forest with Bootstrap  & {\small\cite{breiman2001random}}           & \texttt{randomForest} \\
\emu{confrf} & Conformal Random Forest       & {\small\cite{johansson2014regression}}         & \texttt{conforest} \\
\emu{ngboost}& Natural Gradient Boosting     & {\small\cite{duan2020ngboost}}              & \begin{tabular}{@{}l@{}}
\texttt{Akai01} \\
\ \ \texttt{/ngboost}$^\star$
\end{tabular} \\
\emu{bcart}  & Bayesian CART                 & {\small\cite{chipman1998bayesian}}           & \texttt{tgp} \\
\emu{btreelm}& Treed Bayesian Linear Model   & {\small\cite{chipman2002bayesian}}           & \texttt{tgp} \\
\emu{blasso} & Bayesian LASSO                & {\small\cite{park2008bayesian}}             & \texttt{BayesianLasso} \\
\hline
\end{tabular}
\caption{Summary of emulators and surrogate models compared in this study. Packages available on CRAN are listed by name. For methods not available on CRAN, a GitHub repository is indicated by \texttt{username/repo} or simply \texttt{/repo} if hosted at \url{https://github.com/knrumsey}. The symbol $^\star$ denotes implementations that rely on external Python libraries, accessed via the \texttt{reticulate} package, rather than native R packages \citep{reticulate}. In several cases, dedicated repositories were created to provide a convenient interface to methods that are otherwise unavailable in R.}
\label{tab:emulators}
\end{table}

Deciding on the $32$ emulators to present in this paper was no small task. We sought to create a diverse set of approaches used across many of the various specialized UQ communities, with an emphasis on methods that provide full predictive distributions rather than point estimates. Accordingly, our primary performance metric is continuous ranked probability score (CRPS; see \cref{eq:crps}), a proper scoring rule for probabilistic predictions that directly evaluates this distributional output. Inevitably, many strong alternatives were left out; one has to draw the line somewhere. 

The Gaussian process (GP) is considered by many to be the ``gold-standard" in surrogate modeling, and is thus a good place to start (along with some popular variants). Because GP computation scales cubically with the size of the training data, GPs can become impractical for even moderately sized datasets. For this reason, approximate Gaussian process regression has been a vibrant area of research over the past two decades (see \cite{liu2020gaussian} for an extensive review), and so we include a diverse set of representatives here. Basis function regressors are a broad and popular alternative to GPs (though the GP itself can technically be included in this class), including linear models \citep{lindley1972bayes}, polynomial chaos expansions \citep{wiener1938homogeneous}, multivariate adaptive regression splines \citep{friedman1991}, and many more \citep{denison2002bayesian}. Finally, we include many methods popular in the machine learning community. Although we make an attempt to categorize the results below, we note that these groupings are intended as a useful heuristic and not a definitive classification. Most methods described below could fit into multiple categories. 

Uncertainty estimation is handled in a variety of ways throughout this study. Most commonly, this is done by equipping a method with a probability model and using statistical inference to quantify uncertainty. Bayesian methods are often preferred, both for their well-calibrated uncertainty and their reduced sensitivity to tuning parameter choices. Alongside these, we also include methods based on variational inference, conformal inference and the bootstrap, which offer alternative perspectives on uncertainty quantification.

This study is not intended to provide {\it de facto} answers about which emulator is best. There is no free lunch (see \cref{sec:discuss} for more), and the search for a one-size-fits-all approach is misguided. Many excellent emulation approaches are not considered here, and it is not possible to test the full space of possible computer models (despite our best efforts). For this reason, we evaluate each method using its ``out-of-the-box" settings when possible, making sensible choices (balancing speed and accuracy) based on the literature and software documentation when necessary. 

It is important to note that, with sufficient tuning, the results for any given method could change drastically. Indeed, many of the emulators we consider are universal approximators, capable of emulating any function arbitrarily well given enough training data. With finite data, however, their inference algorithms can yield varying degrees of over- or under-fitting, which practitioners must try to control through tuning. 

Finally, given the behavioral range of the test functions, even the best implementations can occasionally fail. When this happens, we include a series of fallback models to ensure that every function can still be evaluated. We have made every effort to be reasonable in these choices, but the reported performance of each emulator may still be influenced by how often it fails. The R code used to test each method is provided in the supplement. The full list of emulators is given in \cref{tab:emulators}, along with a reference and the associated software package. 

\subsection{Gaussian Processes and Their Variants}

\paragraph{Gaussian Process (\emu{gp}, \emu{rgasp}, \emu{gpytorch}):}
Gaussian processes have long been at the forefront of computer model emulation. In fact, some authors (historically) have used the terms GP and emulator interchangeably. See \cite{rasmussen2006gaussian} or \cite{gramacy2020surrogates} (or with high probability, a randomly selected reference in this paper) for details. 

Though there are many options for Gaussian process modeling in \textsf{R}, we include two state-of-the-art implementations. The first (\emu{gp}) is the \texttt{mleHomGP()} function from the \texttt{hetGP} package \citep{binois2021hetgp}, which is known for its speed and stability. The second (\emu{rgasp}) is the \texttt{rgasp()} function from the \texttt{RobustGaSP} package \citep{gu2018robustgasp}, which offers a sophisticated framework for Gaussian process parameter estimation, along with strong theoretical guarantees \cite{gu2018robust}, but is comparatively less optimized for speed.

We also include a Python-based implementation (\emu{gpytorch}) using the \texttt{gpytorch} library (accessed through \texttt{reticulate}) which provides a modern framework for scalable Gaussian process modeling. In our experiments, this implementation emphasizes computational efficiency but tends to sacrifice some predictive accuracy relative to the R-based implementations \citep{gardner2018gpytorch, reticulate}. All three implementations estimate the nugget, rather than fixing it at zero, so that models can appropriately adapt to non-deterministic settings; for \emu{rgasp}, we set \texttt{nugget.est = TRUE}.

\paragraph{Deep GP (\emu{deepgp}):}
Deep GPs (DGPs), first introduced by \cite{damianou2013deep}, are better equipped to handle nonstationarity and complex response surfaces by stacking multiple latent GP layers, enabling automatic adaptation to regime changes and heterogeneity in the data. We use the \texttt{deepgp} R package to fit a two-layer DGP, leveraging elliptical slice sampling and a Vecchia approximation for efficient Bayesian inference \citep{sauer2023active, sauer2023vecchia}.

\paragraph{Treed GP (\emu{treegp}):}
Treed Gaussian processes (TGPs), introduced by \cite{gramacy2008bayesian}, partition the input space with a tree structure and fit independent GPs within each region. TGPs are particularly useful for emulating computer models that exhibit discontinuities, abrupt regime shifts, or other forms of nonstationarity. We use the \texttt{GPTreeO} R package, which provides an efficient implementation of a TGP surrogate \citep{braun2024gptreeo}.

\paragraph{Heteroskedastic GP (\emu{hetgp}):}
Heteroskedastic Gaussian processes (HetGPs) extend traditional GPs by modeling not only the mean response but also input-dependent noise variance \citep{binois2018practical}. This enables accurate uncertainty quantification even when the data exhibit varying levels of noise across the input space. We use the \texttt{hetGP::mleHetGP()} function, which offers fast and reliable estimation \citep{binois2021hetgp}.

\subsection{Approximate Gaussian Processes}

\paragraph{Matching Pursuit (\emu{mpgp}):}
One simple version of approximate GP regression is known as a subset of data (SoD) approach, in which training samples are discarded until the number of remaining samples is feasible for likelihood evaluation. Selecting a subset of points uniformly at random is typically cheap but ineffective. The Matching Pursuit GP (MPGP) of \cite{keerthi2005matching} constructs the subset sequentially, greedily adding a data point that leads to a large increase in the log posterior of the full GP model (see also \cite{smola2000sparse}). 

We use the implementation in the \texttt{spareGParts} R package (\url{github.com/knrumsey/spareGParts}). Because the MPGP algorithm requires fixed kernel hyperparameters, we apply a simple two-step loop: alternating between subset selection and kernel re-estimation. By default, this loop runs twice before returning the final model.

\paragraph{Local Approximate GP (\emu{lagp}):}
The local approximate GP (laGP) is similar to a SoD approach, except that the sub-design is custom-built for each new prediction point \citep{gramacy2015local}. This allows for good performance with a much smaller GP compared to global SoD approaches and naturally accounts for nonstationarity in the computer model response. Neighborhood subsets are greedily selected to minimize an empirical Bayesian mean square prediction error, improving over the simpler nearest neighbor construction \citep{cressie1990} without drastically increasing the complexity. We use the \texttt{laGP} R package with neighborhood sizes equal to $\min(\max(30, \lfloor\sqrt{n}\rfloor), 100)$ after initializing neighborhoods with the $10$ nearest neighbors \citep{gramacy2016lagp}. As shown in \cref{sec:tune}, \emu{lagp} can be sensitive to the nugget specification. We initialize the nugget at \texttt{g} $= \max(10^{-7}\text{Var}(y), 10^{-9})$.

\paragraph{Active Learning Cohn (\emu{alcgp}):}
Another global SoD approach using the \texttt{laGP} package. This functionality (described as ``beta" in the documentation) uses a similar active learning strategy to construct a single neighborhood for all prediction points jointly. We use a subset size of $\min(\max(100, 2\lfloor\sqrt{n}\rfloor), 300, n-1)$. 

\paragraph{Bayesian Committee Machine (\emu{bcmgp}):}
The Bayesian Committee Machine (BCM) of \cite{tresp2000bayesian} makes GP regressions scalable by partitioning the data into a collection of manageable subsets. A GP, called an ``expert", is fit to each subset, sharing hyperparameters by maximizing the product of individual likelihoods. Unlike in SoD, all of the experts are aggregated during prediction. Our implementation in \texttt{spareGParts} uses the robust BCM of \cite{deisenroth2015distributed}, which weights the experts based on how far they are from the query location. A fast partitioning around medoids (PAM) algorithm is used to choose the $\lfloor \sqrt{n}/2\rfloor$ data partitions. 

\paragraph{Random Fourier Features (\emu{rffgp}):}
The random features approach of \cite{rahimi2007random} maps the data into a randomized low-dimensional feature space, where inner products approximate those of a shift-invariant kernel. This can also be viewed as a sparse basis function regression model. We use the \texttt{gplite} R package with $\min(512, 2\lfloor\sqrt n\rfloor)$ basis functions.

\paragraph{Inducing Point GP (\emu{fitcgp}, \emu{svigp}):}
Inducing point methods are a widely used class of scalable Gaussian process approximations. Like SoD methods, these approximations use a reduced set of ``inducing points'' (also called pseudo-inputs) to improve efficiency; however, these inducing points need not belong to the training data and can be freely optimized, offering greater flexibility.

The fully independent training conditional (FITC) method \citep{snelson2005sparse} is a seminal inducing point approach that defines a sparse approximation to the GP covariance based on the inducing set, and then maximizes the resulting (approximate) marginal likelihood using gradient descent. We use the \texttt{gplite} R package with $\min(\max(100, 2\lfloor\sqrt{n}\rfloor), 300, n-1)$ inducing points.

We also include a stochastic variational inducing point (\emu{svigp}) following \cite{hensman2013gaussian}, which instead introduces a variational distribution over the inducing variables. This formulation leads to an evidence lower bound (ELBO) that decomposes over the original data points, enabling the use of mini-batch stochastic gradient descent for efficient optimization of all hyperparameters. Our implementation uses the \texttt{gpytorch} library in Python, accessed through \texttt{reticulate}.

\paragraph{Scaled Vecchia GP (\emu{svecgp}):}
Vecchia approximations are a powerful tool for approximating complex high-dimensional likelihoods by decomposing the joint density into a sequence of low-dimensional conditional distributions \citep{vecchia1988estimation}. This approach creates a sparse representation of the model leading to tractable inference for large datasets that would otherwise be computationally prohibitive. In the context of GPs, \cite{stein2004approximating} first demonstrated how Vecchia approximations could scale up spatial GP inference and these ideas have been refined and extended by \cite{katzfuss2020vecchia} and \cite{katzfuss2021general}. 

The scaled Vecchia GP used in this paper is described by \cite{katzfuss2022scaled} and is tailored towards computer models by transforming the input space via Fisher scoring. The implementation we use is available in the \texttt{spareGParts} R package, which primarily serves as a wrapper for the \texttt{GpGp} and \texttt{GPvecchia} packages \citep{GpGp, GPvecchia}.

\subsection{Basis Function Regression}

\paragraph{Bayesian Linear Model (\emu{blm}):}
The Bayesian linear model (BLM) serves as a simple reference point, fitting a linear regression with conjugate Bayesian inference for coefficients and noise variance. See, e.g., \cite{lindley1972bayes, gelman2006prior}.

\paragraph{Bayesian Additive Regression Trees (\emu{bart}):}
BART is a popular and widely-used ensemble of regression trees, similar to boosted trees \citep{friedman2001greedy}, but equipped with a probabilistic framework that enables uncertainty quantification \citep{chipman2012bart}. It employs a Bayesian back-fitting MCMC algorithm to sample trees conditionally on the rest of the forest. Prior distributions are placed on tree structure and terminal node parameters, with regularization to control tree depth and prevent overfitting. We use the \texttt{wbart()} function from the \texttt{BART} R package using default settings. 

\paragraph{Bayesian Adaptive Spline Surfaces (\emu{bass}):}
The Multivariate Adaptive Regression Splines (MARS\textsuperscript{\tiny TM}) algorithm was originally proposed by \cite{friedman1991} as a fast and effective way to greedily construct basis functions for spline surfaces from the bottom up (as opposed to top-down or regularization-based approaches). \cite{denison1998} introduced the first Bayesian extension to MARS (BMARS), which was subsequently refined by \cite{nott2005} and \cite{francom2020bass}. BMARS employs a simplified form of reversible jump Markov chain Monte Carlo (RJMCMC; \cite{green1995}) to adaptively learn the number of basis functions, providing principled uncertainty quantification out of the box. We use the \texttt{BASS} R package (named to avoid trademark issues) using default settings.

\paragraph{Generalized BASS (\emu{tbass}, \emu{qbass}):}
Generalized BMARS extends BMARS/BASS by replacing the Gaussian error assumption with the broader class of generalized hyperbolic distributions \citep{rumsey2024generalized}. This framework includes several robust alternatives to standard BASS. We consider two such variants: \texttt{GBASS::tbass()}, which uses a t-distributed error model, and \texttt{GBASS::qbass()}, which targets conditional quantiles. For \emu{tbass}, we use $3$ degrees of freedom; for \emu{qbass}, we set $q=0.5$, corresponding to median regression. The latter induces an inflated Laplace predictive distribution rather than a Gaussian one, which may be advantageous for robust point prediction but less well aligned with scoring rules such as CRPS in the simulation settings considered here. 

\paragraph{Bayesian Projection Pursuit Regression (\emu{bppr}):}
Projection pursuit regression (PPR; \cite{friedman1981projection}) approximates complex multivariate functions as sums of flexible ridge functions applied to linear projections of the inputs, with single index models as a simplified case \citep{ichimura1987estimation}.  The Bayesian version, recently developed by \cite{collins2024bayesian}, provides UQ and uses RJMCMC to infer both the number and structure of the ridge functions. We use the \texttt{BayesPPR} R package (\url{github.com/gqcollins/BayesPPR}) with default settings.

\paragraph{Sparse Bayesian Polynomial Chaos (\emu{spce}):}
Polynomial chaos expansions (PCE) have a long history as a popular emulator in certain UQ sub-communities \citep{wiener1938homogeneous, xiu2002wiener}. PCE expresses the model output as a sum of tensor products of orthogonal polynomials. The approach of \cite{shao2017bayesian} performs efficient Bayesian inference on sparse PCE models with fixed basis functions. Model selection is performed via a ranking procedure coupled with an information criterion. Initially the model is heavily constrained, and the set of candidate basis functions is enriched until a sufficiently expressive model can be found. The \texttt{khaos} R package provides an implementation of the \cite{shao2017bayesian} algorithm with a few minor deviations, including a different enrichment strategy \citep{rumsey2026bayesian}.

\paragraph{Adaptive Bayesian Polynomial Chaos (\emu{apce}):}
Adaptive Bayesian PCE employs a modified BASS algorithm \citep{francom2020bass} with tensor-product Legendre polynomials as the basis and a new prior/proposal to better learn input interactions \citep{rumsey2026bayesian}. We use the implementation in the \texttt{khaos} R package.

\subsection{Machine Learning and Miscellaneous Methods}

\paragraph{Bayesian Neural Network (\emu{bnn}):}
As neural networks and their variants become increasingly popular, it is no surprise that they are being used as emulators for UQ \citep{sun2019review, tripathy2018deep}. Traditional NNs often provide poor UQ, so we focus here on Bayesian neural networks (BNNs), which offer a principled probabilistic framework for UQ \citep{neal2012bayesian}. Our implementation uses the \texttt{bnns} R package \citep{bnns}, which is built on Stan \citep{carpenter2017stan}, and we fit a BNN with two hidden layers with $8$ nodes each. 

\paragraph{Generative Bayesian Computation (\emu{gbc}):}
Generative Bayesian Computation (GBC), recently proposed by \cite{polson2026generative}, is a likelihood-free surrogate modeling approach that builds on ideas from implicit quantile networks (IQNs) \citep{dabney2018implicit}. Rather than specifying an explicit probabilistic model, GBC learns the conditional quantile function of the response given the inputs, allowing predictive samples to be generated by evaluating the model at randomly drawn quantile levels. This yields a flexible, scalable neural network-based surrogate that does not rely on assumptions of Gaussianity or stationarity.  We use the reference implementation available at \url{github.com/VadimSokolov/gbc-surrogate}, interfaced through \texttt{reticulate}. The model is trained using default settings, and predictive samples are obtained by repeatedly sampling from the learned quantile function.

\paragraph{Relevance Vector Machine (\emu{rvm}):}

The relevance vector machine (RVM), proposed by \cite{tipping1999relevance}, is designed as a probabilistic extension of the support vector machine \citep{cortes1995support}, with the added advantage of typically producing sparser representations. RVMs build predictions as weighted sums of kernel functions, but use (empirical) Bayesian inference to account for uncertainty. Our implementation follows Tipping’s original formulation, with an added discrete prior over the lengthscale parameter to guard against poor kernel choices. The code is available in the \texttt{spareGParts} R package.

\paragraph{Bootstrapped Random Forest (\emu{bootrf}):}
Random forests (RF) are a popular machine learning (ML) technique \citep{breiman2001random}, and their strong predictive performance makes them an obvious candidate for inclusion in the present work. However, standard RFs do not directly provide UQ, so special care must be taken (see also the discussion on BART). It is sometimes (dubiously) claimed that the spread of the weak-learner predictions in boosting ensembles characterizes uncertainty in a meaningful way; we take an approach here which is similar in spirit, but slightly more principled. Specifically, we train a ``super-forest" of $100$ RFs using the nonparametric bootstrap \citep{efron2000bootstrap}, resampling the data for each fit. The collection of predictions from these models is then interpreted (for better or for worse) as a predictive distribution. We use the \texttt{randomForest} R package for implementation. 

\paragraph{Conformal Random Forest (\emu{confrf}):}
Conformal inference (CI) provides a principled framework for constructing valid predictive intervals, ensuring coverage at any predefined confidence level \citep{vovk2005algorithmic}. CI will be used here to provide UQ for random forests. Specifically, we use the RFoK framework discussed by \cite{johansson2014regression}, which efficiently leverages out-of-bag estimates from the random forest for conformal calibration, removing the need for a separate validation set. Our implementation is in the \texttt{conforest} R package (\url{github.com/knrumsey/conforest}) which uses \texttt{randomForest} under the hood.

\paragraph{Natural Gradient Boosting (\emu{ngboost}):}
Natural Gradient Boosting (NGBoost) extends gradient boosting machines, like XGBoost \citep{chen2016xgboost}, to allow for probabilistic predictions. By using natural gradients and flexible multiparameter loss functions, NGBoost can fit a wide variety of outcome distributions and provides meaningful UQ \citep{duan2020ngboost}. It can be paired with any base learner and any continuous parametric family; in this study, we opt for a Gaussian distribution and decision trees as the base learner. We use the R package located at \texttt{Akai01/ngboost} which calls the Python \texttt{ngboost} library via \texttt{reticulate} \citep{reticulate}.

\paragraph{Bayesian CART (\emu{bcart}):}
Bayesian CART (BCART) is a fully Bayesian version of the classic Classification and Regression Trees (CART) algorithm of \cite{breiman2017classification}, enabling probabilistic uncertainty quantification in tree-based models. Two foundational BCART algorithms were proposed independently in 1998 \citep{chipman1998bayesian, denison1998bcart}, each introducing a Bayesian framework for recursive partitioning and inference on trees. We use the implementation from the \texttt{tgp} R package, which follows the approach of \cite{chipman1998bayesian}.

\paragraph{Bayesian Treed Linear Model (\emu{btreelm}):}
The Bayesian treed linear model (BTREELM), introduced in \cite{chipman2002bayesian}, extends the BCART framework by placing a linear model at each of the terminal nodes rather than a constant mean. This hybrid approach combines the interpretability of trees with the flexibility of local linear modeling, often improving predictive accuracy while retaining coherent Bayesian inference. It is also implemented in the \texttt{tgp} package.

\paragraph{Bayesian LASSO (\emu{blasso}):}
The Bayesian LASSO introduces a Laplace prior on regression coefficients, yielding a fully Bayesian extension of the widely used LASSO regularization method \citep{tibshirani1996regression}. The Bayesian formulation of \cite{park2008bayesian} enables rigorous uncertainty quantification and typically requires less tuning of penalty parameters. We use the implementation in the \texttt{BayesianLasso} R package \citep{BayesianLasso}.

\paragraph{Baseline Model (\emu{baseline}):}
This model estimates the mean and variance of the response variable (independent of the inputs) and makes predictions from the resulting Gaussian distribution. This model is used as a baseline for comparison to identify scenarios where more sophisticated emulators are prone to overfitting.

\section{The \texttt{duqling} Package}
\label{sec:duqling}

Fair and reproducible comparison of emulation methods across different settings is challenging. The \texttt{duqling} framework was developed to address this need by providing a standardized and extensible platform for running large-scale emulator comparisons, enabling results to be meaningfully compared, even across papers. 

To illustrate how the framework works in practice, the following R code snippet shows how a user could compare a new emulator against those used in this paper. While the details of the simulation study will be introduced in the next section, the example below gives a preview of how easily such comparisons can be made: 
\begin{rcode}
    library(duqling)

    # Run simulation study
    my_results <- run_sim_study(
                     fit_shiny_new_emulator,   # User-defined
                     pred_shiny_new_emulator,  # User-defined
                     fnames=get_paper_funcs(), # List of test functions
                     n_train=1000,             # Training set size
                     NSR=c(0, 0.1),            # Noise-to-signal ratio
                     design_type="LHS",        # Design matrix generation
                     replications=10,          # Number of replications 
                     mc_cores=10)              # Number of cores to use
\end{rcode}
The function \texttt{run\_sim\_study()} executes a full simulation study by applying a user-defined emulator to a configurable set of test functions, design settings, noise levels, and replications. The design type can be set to (i) Latin hypercube sampling (LHS; \citep{mckay1979comparison}), (ii) independent uniform sampling, (iii) a uniform grid, or (iv) user-specified custom designs. The noise-to-signal ratio (NSR) is defined as the ratio of the observational noise variance to the unconditional variance of the simulator response, i.e., $NSR = \mathrm{Var}(\epsilon)/\mathrm{Var}(f(\bm x))$.

After running the simulation study, any of the figures in \cref{sec:results} can be created with just a few lines of code, as shown below.
\begin{rcode}
    # Load the paper data
    data("sim_study_testfuncs")

    # Process and combine simulation studies
    duq_paper <- process_sim_study(sim_study_testfuncs)
    duq_myres <- process_sim_study(my_results)
    duq_joined <- join_sim_study(duq_paper, duq_myres) 
    
    # Filter for specific case
    duq_joined_filtered <- filter_sim_study(duq_joined, n_train=1000, NSR=0)
    
    # Make figures and analysis
    summarize_sim_study(duq_joined_filtered)
    rankplot_sim_study(duq_joined_filtered, metric="CRPS")
\end{rcode}

A Python version of the package is also available at \url{https://github.com/reidmorris/duqling_py}, though it is less actively maintained than the R implementation. Alternatively, the R package can be accessed from Python using interoperability tools such as \texttt{rpy2} \citep{rpy2}.

\subsection{Test Functions}
The \texttt{duqling} package includes a large (and growing) collection of test functions. This paper focuses on scalar-output, deterministic functions with continuous inputs, and the automated simulation study framework currently supports only this setting. The package also includes test functions representing stochastic simulators, functions with categorical inputs, and multivariate or functional outputs, although automated simulation studies for these cases are still in development.

The set of test functions used here was curated to reflect a broad range of modeling challenges. These include functions with input dimensions ranging from $1$ to $100$; both stationary and nonstationary behavior; smooth and discontinuous surfaces; low and high effective dimension; and varying degrees of linearity and nonlinearity. Several test functions include fully inert variables or are constant-valued, which helps test for overfitting.  

All $60$ test functions are listed in a table in the supplement, along with the input dimension and some statistical summaries. References and additional details for each function are provided in the \texttt{duqling} package documentation; many are also described in greater detail in the Simon Fraser virtual library of simulation experiments \citep{surjanovic2013virtual}.

\begin{table}[!htbp] \centering 
  \caption{Summary of real-world datasets used in this study. The number of observations is $n$ and the number of inputs is $p$.} 
  \label{tab:datasets} 
\begin{tabular}{@{\extracolsep{5pt}} lcccl} 
\\[-1.8ex]\hline 
\hline \\[-1.8ex] 
Dataset & $n$ & $p$ & \texttt{duqling} & Reference \\ 
\hline \\[-1.8ex] 
\tf{plate\_deformation} & $138$ & $7$ & No & {\scriptsize \cite{francom2018sensitivity}} \\ 
\tf{wind\_speed} & $200$ & $17$ & No & {\scriptsize \cite{edmunds2013value}} \\ 
\tf{strontium\_plume\_p4b} & $300$ & $20$ & Yes & {\scriptsize \cite{volkova2008global}} \\ 
\tf{strontium\_plume\_p104} & $300$ & $20$ & Yes & {\scriptsize \cite{volkova2008global}} \\ 
\tf{spectra1} & $500$ & $4$ & No & {\scriptsize \cite{klein_2020_4069493}} \\ 
\tf{spectra2} & $500$ & $4$ & No & {\scriptsize \cite{klein_2020_4069493}} \\ 
\tf{pbx9501\_gold} & $500$ & $6$ & Yes & {\scriptsize \cite{rumsey2025coactive}} \\ 
\tf{pbx9501\_ss304} & $500$ & $6$ & Yes & {\scriptsize \cite{rumsey2025coactive}} \\ 
\tf{pbx9501\_nickel} & $500$ & $6$ & Yes & {\scriptsize \cite{rumsey2025coactive}} \\ 
\tf{pbx9501\_uranium} & $500$ & $6$ & Yes & {\scriptsize \cite{rumsey2025coactive}} \\ 
\tf{ptw1} & $500$ & $10$ & No & {\scriptsize \cite{impala}} \\ 
\tf{discoflux\_flyer} & $914$ & $6$ & No & {\scriptsize \cite{francom2021simulation}} \\ 
\tf{taylor\_foot} & $962$ & $11$ & No &  {\scriptsize \cite{biswas2021asc}}\\ 
\tf{taylor\_length} & $962$ & $11$ & No & {\scriptsize \cite{biswas2021asc}} \\ 
\tf{jc} & $1{,}000$ & $5$ & No & {\scriptsize \cite{impala}}  \\ 
\tf{flyerPTW1} & $1{,}000$ & $11$ & No &  {\scriptsize \cite{biswas2021asc}}\\ 
\tf{flyerPTW2} & $1{,}000$ & $11$ & No &  {\scriptsize \cite{biswas2021asc}}\\ 
\tf{flyer\_plate104} & $1{,}000$ & $11$ & Yes & {\scriptsize \cite{walters2018bayesian}} \\ 
\tf{diffusion\_1D} & $1{,}000$ & $62$ & No & {\scriptsize \cite{hlobilova_2024_12704504}} \\ 
\tf{fair\_climate\_ssp1-2.6\_year2200} & $1{,}001$ & $45$ & Yes & {\scriptsize \cite{smith2018fair}} \\ 
\tf{fair\_climate\_ssp2-4.5\_year2200} & $1{,}001$ & $45$ & Yes &  {\scriptsize \cite{smith2018fair}} \\ 
\tf{fair\_climate\_ssp3-7.0\_year2200} & $1{,}001$ & $45$ & Yes &  {\scriptsize \cite{smith2018fair}} \\ 
\tf{taylor\_cylinder1} & $1{,}438$ & $8$ & No &  {\scriptsize \cite{sjue2023simple}}\\ 
\tf{stochastic\_sir} & $2{,}000$ & $4$ & Yes & {\scriptsize \cite{rumsey2024generalized}} \\ 
\tf{diffusion\_2D} & $2{,}000$ & $53$ & No & {\scriptsize \cite{hlobilova_2024_12701147}}\\ 
\tf{acme\_climate} & $2{,}980$ & $66$ & No & {\scriptsize \cite{sargsyan2015sparse}} \\ 
\tf{ptw2} & $3{,}701$ & $13$ & No & {\scriptsize \cite{impala}}  \\ 
\tf{SLOSH\_low} & $4{,}000$ & $5$ & Yes & {\scriptsize \cite{hutchings2023comparing}} \\ 
\tf{SLOSH\_mid} & $4{,}000$ & $5$ & Yes & {\scriptsize \cite{hutchings2023comparing}} \\ 
\tf{SLOSH\_high} & $4{,}000$ & $5$ & Yes & {\scriptsize \cite{hutchings2023comparing}} \\ 
\tf{Z\_machine\_max\_vel1} & $5{,}000$ & $6$ & Yes & {\scriptsize \cite{brown2018estimating}} \\ 
\tf{Z\_machine\_max\_vel2} & $5{,}000$ & $6$ & Yes & {\scriptsize \cite{brown2018estimating}} \\ 
\tf{Z\_machine\_max\_vel\_all} & $5{,}000$ & $30$ & Yes & {\scriptsize \cite{brown2018estimating}} \\ 
\tf{e3sm\_mnar} & $9{,}122$ & $2$ & Yes &  {\scriptsize \cite{grosskopf2021situ}} \\ 
\tf{nuclear\_data} & $9{,}307$ & $178$ & No & {\scriptsize \cite{francom2019nuclear}} \\ 
\tf{e3sm\_mcar} & $10{,}000$ & $2$ & Yes & {\scriptsize \cite{grosskopf2021situ}} \\ 
\tf{icf1} & $10{,}000$ & $5$ & No & {\scriptsize \cite{JAG_LLNL}} \\ 
\tf{icf2} & $10{,}000$ & $5$ & No & {\scriptsize \cite{JAG_LLNL}} \\ 
\tf{diablo\_canyon\_plume} & $18{,}000$ & $13$ & No & {\scriptsize \cite{francom2019inferring}} \\ 
\tf{okc\_plume} & $22{,}229$ & $5$ & No & {\scriptsize \cite{brown2009evaluation}} \\ 
\hline \\[-1.8ex] 
\end{tabular} 
\end{table}

\subsection{Data Sets}
We also evaluate emulator performance on 40 real-world datasets, $19$ of which are publicly available in the \texttt{duqling} package, with the remaining $21$ not yet approved for release. The datasets span a wide range of applications, including problems in chemistry, materials science, nuclear physics, climate modeling, epidemiology, and plume dispersion. Sample sizes range from 138 to $22{,}229$, and input dimensionality ranges from $2$ to $178$. In some cases, inputs were selected using formal design of experiments, while in others they represent posterior draws from a calibration procedure \citep{kennedy2001bayesian} and may exhibit strong correlation.

For analysis and visualization, we divide the datasets into two groups based on training size: small datasets with $n < 2000$ and large datasets with $n \geq 2000$. A full list of datasets, along with their sample sizes, input dimensions, and references, is provided in \cref{tab:datasets}.

\subsection{Reproducible Simulation Studies}
A central design goal of \texttt{duqling::run\_sim\_study()} is strict reproducibility. For any combination of test function, training size, design type, noise level, and replication index, the generated training data—both inputs and responses—are deterministic. This holds even when simulations are run in different orders or batches. For example:
\begin{rcode}
    emulators <- get_emulator_functions(c("rgasp", "lagp", "bart"))
    results1 <- run_sim_study(emulators$fit_func, emulators$pred_func,
                              fnames = c("borehole", "ishigami"),
                              NSR = 0, 
                              replications = 1:10)

    results2 <- run_sim_study(emulators$fit_func, emulators$pred_func,
                              fnames = "ishigami",
                              NSR = 0, 
                              replications = c(1, 7))
\end{rcode}
The rows corresponding to \texttt{ishigami}, \texttt{NSR = 0}, and the specified replications will be identical in both cases (except for timing, which is machine-dependent). Internally, each simulation scenario is transformed into a unique seed via a polynomial hashing scheme, ensuring reproducibility across platforms and workflows. 

For real-world datasets, \texttt{run\_sim\_study\_data()} provides analogous functionality, using dataset name, fold number, and cross-validation type (both cross-validation and bootstrap are currently supported) as the hash keys. Full details of the hashing algorithm are provided in the supplement.

\subsubsection{Fallback Model}
\label{sec:fallback}

To maintain consistency across such a large and diverse test suite, \texttt{duqling::run\_sim\_study()} includes a fallback model that activates when an emulator fails during fitting or prediction. The fallback is a simple null model that predicts from a Student’s $t$ distribution fit to the marginal distribution of the training responses. While deliberately naive, it ensures that performance metrics remain well-defined even when an emulator breaks. Failures are recorded in the output via a categorical \texttt{failure\_type} field, which takes the values \texttt{"none"}, \texttt{"fit"}, or \texttt{"pred"}. 

The continuous ranked probability score (CRPS) serves as our primary performance metric (see \cref{eq:crps}). For a unit-variance response, the baseline model has an expected CRPS of $\pi^{-1/2} \approx 0.564$, which we use as a reference point after rescaling. Additional details are provided in \cref{sec:results}.

\subsubsection{Simulation Output Format}

The function \texttt{run\_sim\_study()} returns an R data frame where each row corresponds to a single emulator applied to a specific simulation scenario and replication. Key columns include:
\begin{itemize}
    \item \texttt{method}, \texttt{fname}, \texttt{n\_train}, \texttt{NSR}, \texttt{design\_type}, and \texttt{replication}, identifying the simulation configuration.
    \item \texttt{RMSE}, \texttt{FVU}, and \texttt{CRPS} as primary performance metrics, where \texttt{RMSE} is root mean squared error, \texttt{FVU} is the fraction of variance unexplained (see Section SM5.1 for definition), and \texttt{CRPS} is the continuous ranked probability score (see \cref{eq:crps}).
    \item \texttt{t\_fit}, \texttt{t\_pred}, and \texttt{t\_tot}, for fit time, prediction time, and total runtime in seconds, respectively.
\end{itemize}

Additional columns can optionally be included when we seek to further assess predictive uncertainty, including empirical coverage (\texttt{COVER<level>}), mean interval length (\texttt{LENGTH<level>}), mean interval score (\texttt{MIS<level>}), and summary statistics of CRPS across test points. A complete description is provided in the supplement.

When $NSR>0$, \texttt{duqling} distinguishes between latent and noisy test evaluations. By default, point-prediction metrics (RMSE, MAE, Bias, and FVU) are computed against the latent simulator response, whereas distributional and interval-based metrics (including CRPS and coverage) are computed against the noisy test response.

The function \texttt{run\_sim\_study\_data()} returns results in the same format, but replaces \texttt{fname} and \texttt{replication} with \texttt{dname}, \texttt{fold}, and \texttt{fold\_size}, reflecting its use of cross-validation. The full simulation results used in this paper are bundled with the \texttt{duqling} package and can be accessed via \texttt{data(sim\_study\_testfuncs)} and \texttt{data(sim\_study\_realdata)}.

\begin{figure}[ht]
    \centering
 \includegraphics[alt={Cumulative rank curves showing strong performance by several Gaussian process and basis-function emulators. See caption for details.}, width=0.98\linewidth]{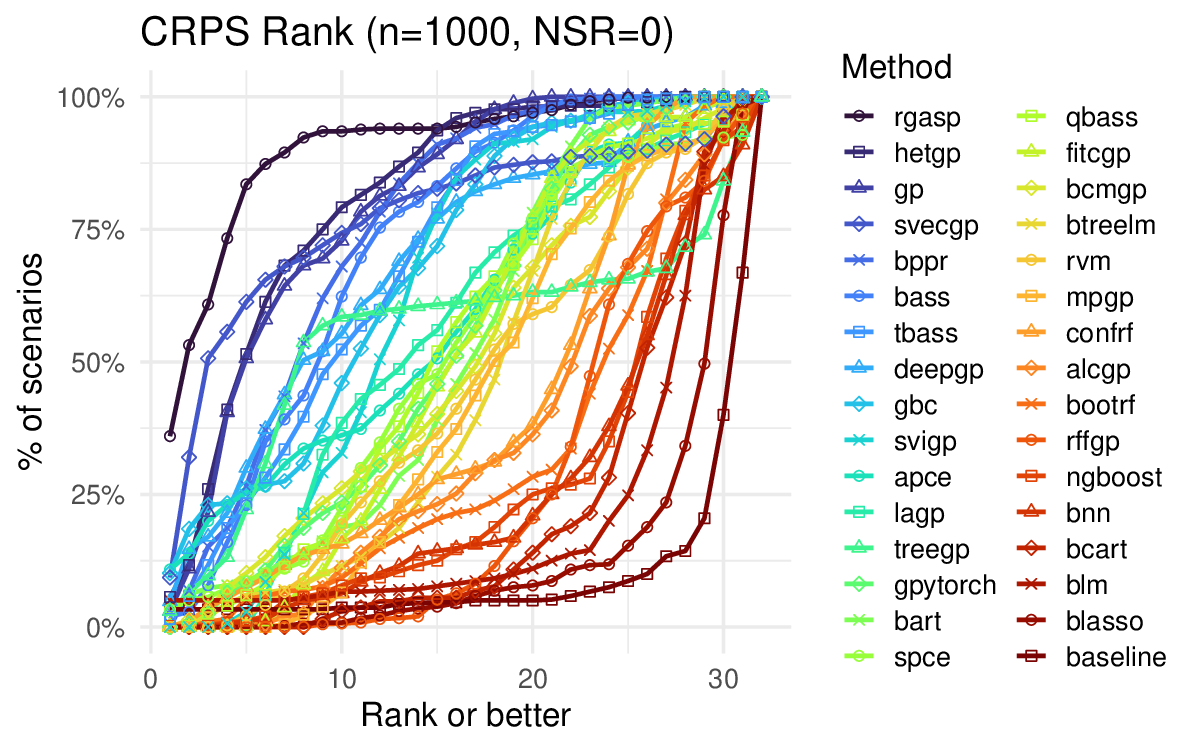}
    \caption{Cumulative rank-plot for $n=1000$, $NSR=0$ setting. The curve for each emulator shows the proportion of cases that the method was at least top $r$ out of $32$ in terms of CRPS, for $r=1,\ldots, 32$. }
    \label{fig:rank_1a}
\end{figure}

\begin{figure}[ht]
    \centering    \includegraphics[alt={Cumulative rank curves showing several competitive emulators and shifts in relative performance with added noise. See caption for details.  }, width=0.98\linewidth]{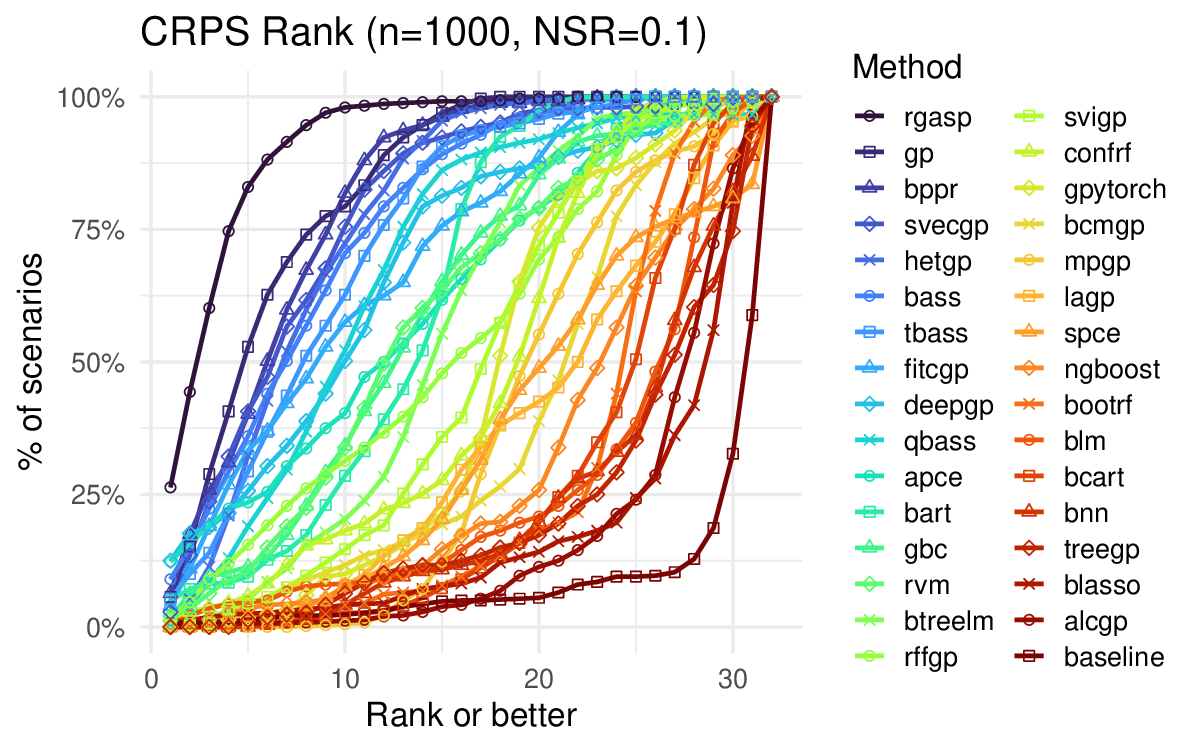}
    \caption{Cumulative rank-plot for $n=1000$, $NSR=0.1$ setting. The curve for each emulator shows the proportion of cases that the method was at least top $r$ out of $32$ in terms of CRPS, for $r=1,\ldots, 32$. }
    \label{fig:rank_1b}
\end{figure}

\begin{figure}[h]
    \centering    \includegraphics[alt={Heatmap showing substantial variation in emulator performance across test functions, with no method uniformly best. See caption for details.  }, width=0.98\linewidth]{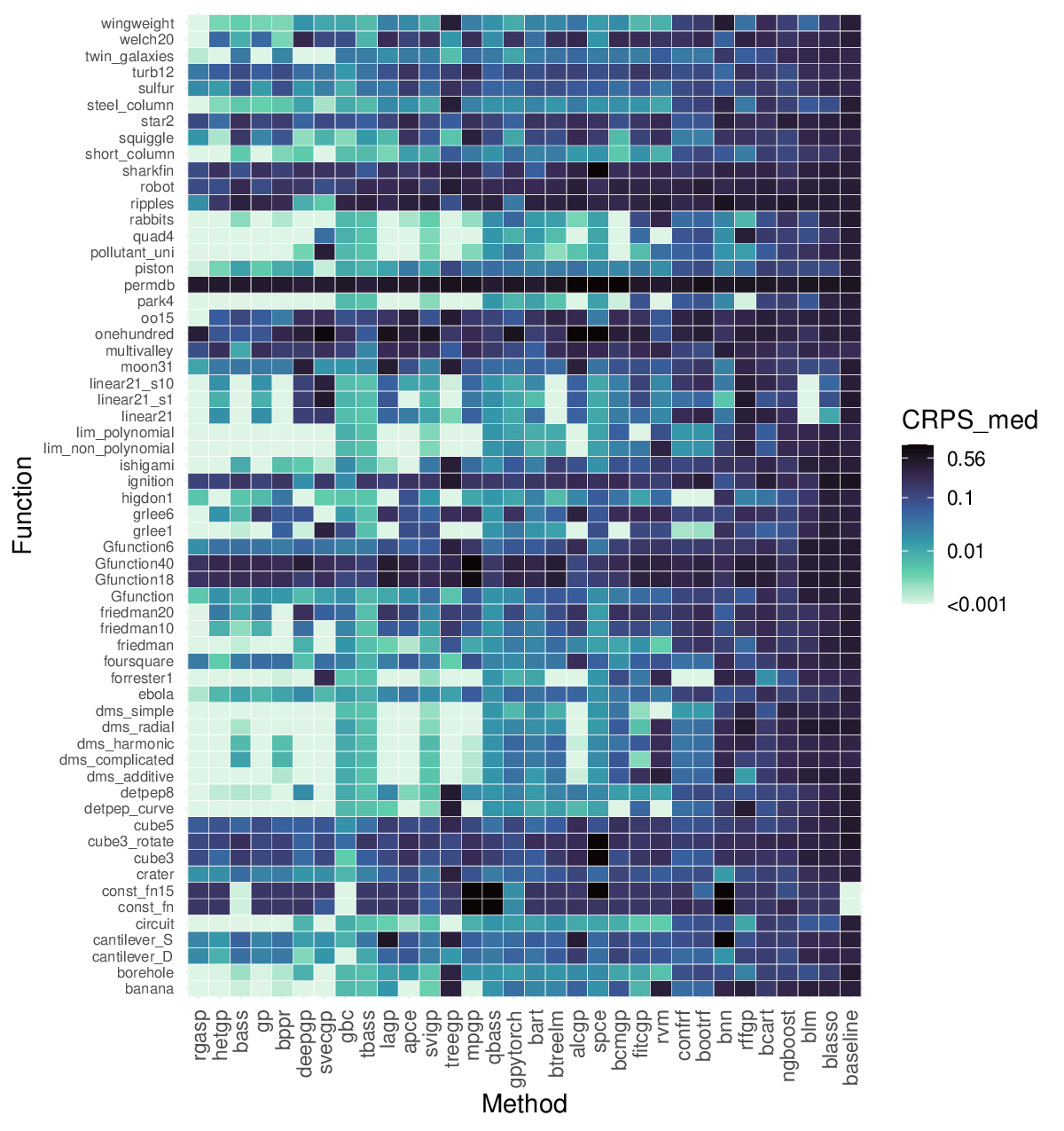}
    \caption{A heatmap of CRPS (truncated between $0.001$ and $1.0$ for visual simplicity) for the synthetic test functions in the $n=1000, NSR=0$ setting.}
    \label{fig:heat_1a}
\end{figure}

\section{Results}
\label{sec:results}

In this section, we compare the results of $32$ emulators across $60$ synthetic test functions and $40$ real-world simulation datasets. Given the scope of the study, the analysis here is necessarily high-level, but in \cref{sec:discuss} we demonstrate how \texttt{duqling} can be used to conduct more granular, targeted investigations. All performance metrics are computed on out-of-sample test sets. For synthetic test functions, we use maximin Latin hypercube designs with $n_\text{test} = 1000$ to assess out-of-sample performance \citep{mckay1979comparison}. For the real-world datasets, we apply $10$-fold cross-validation, so $n_\text{test}$ is determined by the size of the dataset. Our primary performance metric is continuous ranked probability score (CRPS), a proper scoring rule for probabilistic predictions that evaluates both predictive accuracy and uncertainty quantification simultaneously. CRPS is widely regarded as a gold-standard metric for evaluating predictions made with uncertainty \citep{gneiting2007strictly}. For $i=1,2,\ldots, n_\text{test}$
\begin{equation}
\label{eq:crps}
\text{CRPS}_i = \text{CRPS}(y_{\text{test},i}, \{\hat y_1, \ldots, \hat y_M\}) = \frac{1}{M}\sum_{j=1}^M|\hat y_j - y_{\text{test},i}| - \frac{1}{M(M-1)}\sum_{j<k}|\hat y_j - \hat y_k|.
\end{equation}
where $\hat y_1, \ldots, \hat y_M$ are predictive samples from the emulator for the $i^{th}$ test point.  Metrics based solely on point predictions, such as FVU (see Section SM5.1 for definition), are also available but do not assess uncertainty quantification. Alternative diagnostics, including empirical coverage and interval-based metrics, can also be computed within the \texttt{duqling} framework (see Section SM5.4), but are not the primary focus of this study.

To facilitate interpretation and comparability across datasets, we rescale all CRPS values as if the true response had unit variance. On this scale, a CRPS value of $1/\sqrt{\pi} \approx 0.564$ corresponds to the expected score of the baseline model, which is useful for interpretation. In practice, the difference between sufficiently small CRPS values is largely irrelevant beyond a certain point. This is particularly true in many primary applications of emulators, such as sensitivity analysis, optimization, and calibration. Sensitivity analysis techniques are often robust to small to moderate emulator errors (though overfitting presents more of a challenge). The degree to which calibration and optimization are affected depends on the relative size of measurement and model discrepancy uncertainties. Thus, while many of our analyses report rankings and show differences on this sensitive scale, we note that all CRPS values below a certain threshold (e.g., $0.001$) are likely to be indistinguishable for many applications.

There are many ways to summarize the $n_\text{test}$ CRPS values computed for each emulator on each test set, each highlighting a different aspect of emulator performance. In the analyses that follow, each visualization is based on a different perspective: cumulative rank plots rank emulators based on mean CRPS, heatmaps display median CRPS (over the $n_\text{test}$ test points) averaged across replications, and Pareto front plots compare methods using a relative mean CRPS measure. Specifically, for emulator $i$ and simulation scenario $j$, we define
\begin{equation}
\label{eq:rel_crps}
    t_{i,j} = \frac{\text{CRPS}_{i,j} + \omega}{\min_k\left[ \text{CRPS}_{k,j} \right] + \omega},
\end{equation}
which measures performance relative to the best-performing emulator in each scenario, where $\omega \geq 0$ is a small smoothing parameter, and $\text{CRPS}_{i,j}$ is the mean across the $n_\text{test}$ test points. Choosing a non-zero $\omega$ indicates that values of CRPS much smaller than $\omega$ should all be considered equally good, from a pragmatic perspective. For Pareto front plots, the value $\omega = 0.001$ is used, and the $t_{i,j}$ are capped at $1000$ to reduce the influence of extreme outliers on the overall results. 

For the real-world datasets, we also make use of performance profile plots \citep{dolan2002benchmarking, more2009benchmarking}, which provide an alternative view of relative performance across datasets. The performance profile for emulator $i$ is given by
$$
    \rho_i(\tau) = \frac{1}{J}\sum_{j=1}^J \mathbbm{1}(t_{i,j} \leq \tau), 
$$
which represents the proportion of the $J$ simulation scenarios for which emulator $i$ is within a factor $\tau$ of the best-performing method (we use $\omega = 0$ here). In practice, we find performance profiles particularly useful for the real datasets, where the number of scenarios is smaller and practical differences are of primary interest. For the synthetic test functions, however, the wide range of behaviors and larger number of scenarios make performance profiles less readable and interpretable, so we rely primarily on rank-based summaries in that setting.

With such a large and rich dataset, there are dozens of plausible analyses one could conduct, each with its own variants, and the resulting story can vary significantly depending on the perspective taken. In what follows, we present a range of viewpoints, with many additional analyses and figures provided in the supplement. For the sake of brevity, however, some choices had to be made. We encourage interested readers to explore their own analyses using the tools and data provided. All plots are generated using the \texttt{duqling} package (see supplement for details).

\begin{figure}[h]
    \centering
    \includegraphics[alt={ Heatmap showing substantial variation in emulator performance across noisy test functions, with no method uniformly best. See caption for details }, width=0.98\linewidth]{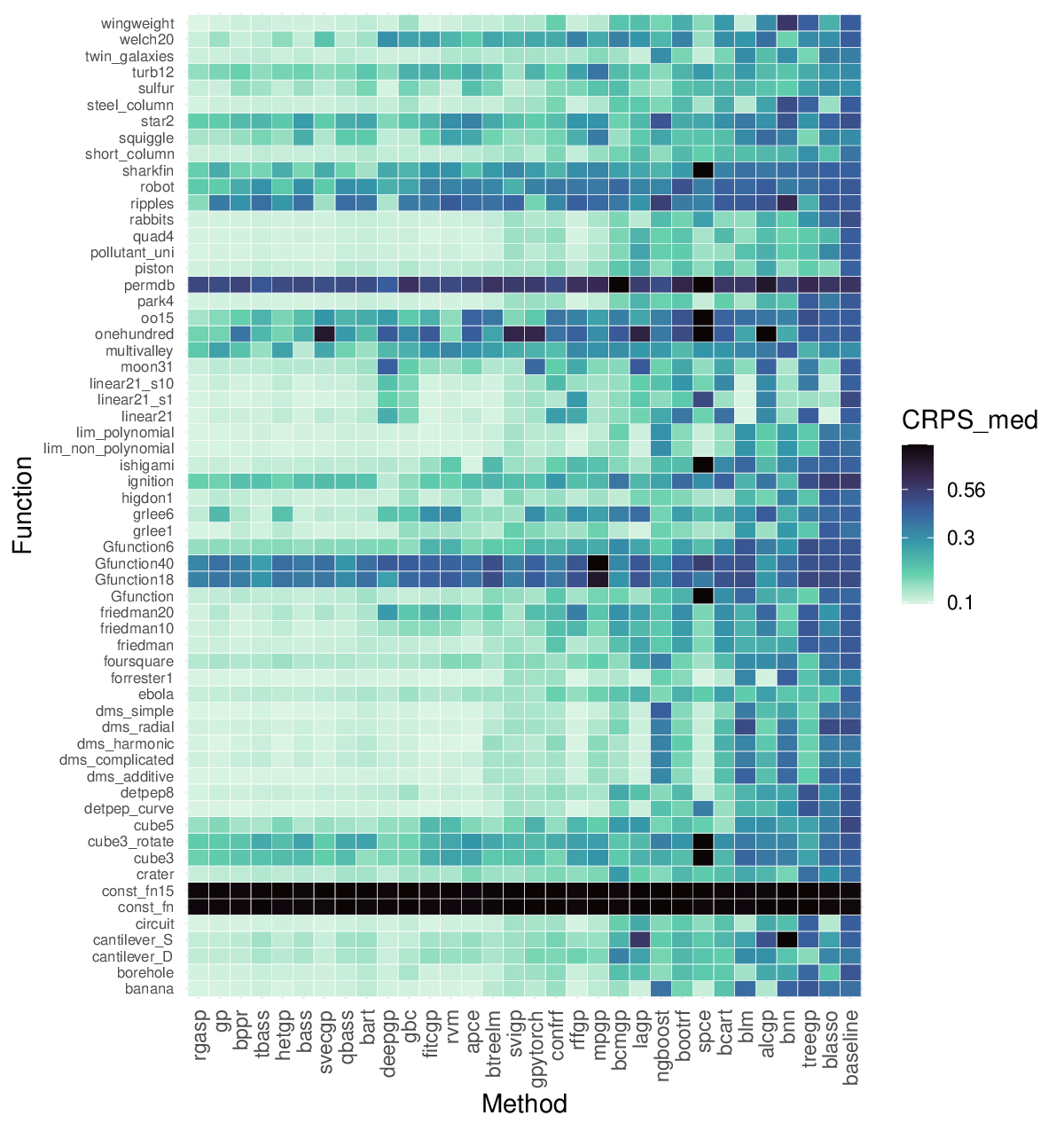}
    \caption{A heatmap of CRPS (truncated between $0.001$ and $1.0$ for visual simplicity) for the synthetic test functions in the $n=1000, NSR=0.1$ setting.}
    \label{fig:heat_1b}
\end{figure}

\paragraph{Simulation Settings.}
We consider the following benchmark settings:
\begin{itemize}
\item $n = 1000$, $NSR = 0$, $10$ replications, $60$ test functions.
\item $n = 1000$, $NSR = 0.1$, $10$ replications, $60$ test functions.
\item $n = 500$, $NSR = 0$, $10$ replications, $60$ test functions. All emulators.
\item $n = 500$, $NSR = 0.1$, $10$ replications, $60$ test functions. All emulators (see supplement).
\item $n = 5000$, NSR = 0, 10 replications, 60 test functions. Excluded: \emu{gp, rgasp, deepgp, hetgp}.
\item 21 real datasets with $n < 2000$. All emulators.
\item 19 real datasets with $n \geq 2000$. Excluded: \emu{gp, rgasp, deepgp, hetgp}.
\end{itemize}
In the computer model emulation setting, the deterministic case ($NSR=0$) is often the primary focus, but setting $NSR=0.1$ allows us to check for overfitting, simulate more complex dynamics (some of which manifest as chaos or near stochastic behavior), or introduce light stochasticity which is sometimes of interest \citep{baker2022analyzing}. 

\paragraph{Caveats and Practical Considerations.}
We emphasize caution in interpreting these results. While we made an effort to use reasonable defaults for all emulators, these results reflect the practical implementation of each emulator as much as the underlying modeling approach. Performance might change considerably with sufficient tuning or different hyperparameter choices. We encourage readers to modify settings using \texttt{duqling} to explore these effects directly.

For computational reasons, emulators were excluded from a given setting if all simulation scenarios took more than 10 hours. 

Across all methods, we use out-of-the-box fitting routines as much as possible and do not assess convergence of any of the algorithms. This is especially relevant for the GPyTorch-based methods, whose hyperparameters are estimated by iterative gradient-based optimization. While this allows these methods to complete in larger-data settings, predictive accuracy may suffer when the optimizer has not fully converged and could potentially improve with additional optimization.

Finally, although \texttt{duqling} separately reports training and prediction time, we report only the total runtime here. In some cases, these phases are not separable due to an emulator’s structure or implementation. This may lead to misleading impressions for some methods (e.g., \emu{svecgp}, \emu{lagp}) that train quickly but predict more slowly, but in other contexts, omitting prediction time could be equally misleading. For example, in applications requiring MCMC (e.g., model calibration) fast prediction is often critically important and cannot be parallelized; see \cite{rumsey2023localized} for discussion. We also note that runtime comparisons should be interpreted with caution, as computational performance can vary substantially across systems. The experiments in this study were run on a Linux-based HPC cluster using 10 cores per task, on nodes with a mix of Intel and AMD processors (x86\_64 architecture). As a result, the timing results reported here are most useful for relative comparisons within this study rather than as absolute benchmarks across papers or computing environments.

\begin{figure}[ht]
  \centering
  \begin{subfigure}[t]{0.48\textwidth}
    \centering
    \includegraphics[alt={Pareto plot showing accuracy and computational-cost tradeoffs in the noise-free setting. See caption for details.}, width=\linewidth]{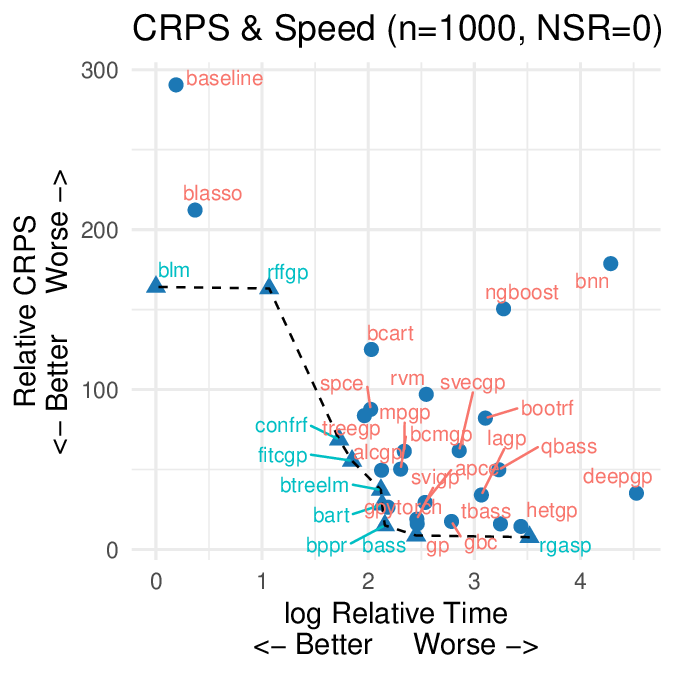}
    \caption{}
    \label{fig:pareto1}
  \end{subfigure}
  \hfill
  \begin{subfigure}[t]{0.48\textwidth}
    \centering
    \includegraphics[alt={Pareto plot showing accuracy and computational-cost tradeoffs in the noisy setting. See caption for details.}, width=\linewidth]{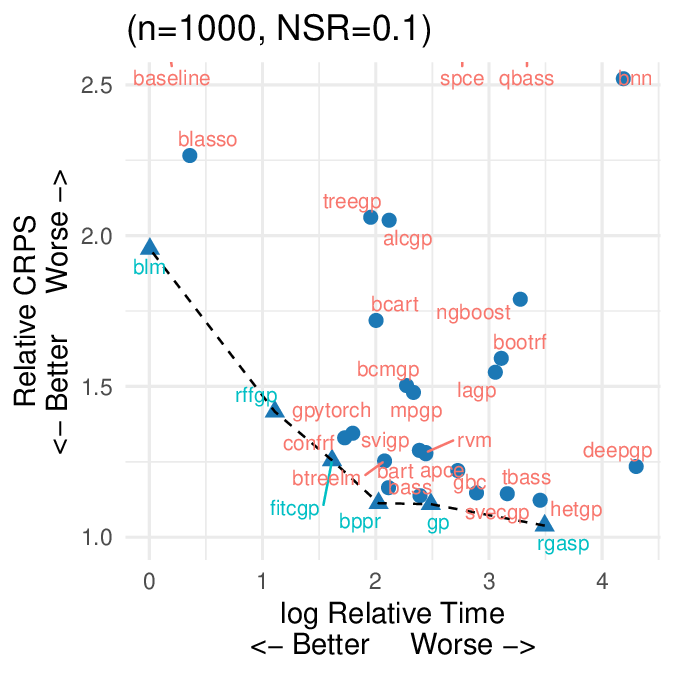}
    \caption{}
    \label{fig:pareto2}
  \end{subfigure}
  \caption{Pareto front of emulators based on average relative CRPS and log relative runtime across all simulation scenarios for the $NSR=0$ and $NSR=0.1$ settings ($n=1000$). Relative CRPS is computed as ($\text{CRPS} + \omega) / (\min \text{CRPS} + \omega$) with $\omega = 0.001$ for each scenario, capped at $1000$ to reduce sensitivity to outliers. The dashed line and triangle markers indicate the methods which are not dominated (slower and less accurate) by any other emulator.}
  \label{fig:combined}
\end{figure}

\subsection{Test Functions (n=1000)}

In our experience, the case where approximately $n=1000$ model evaluations are available for training is both common and interesting. For many computer models, this is a pragmatic upper bound on how many times the true simulator can be run given typical high performance computing (HPC) resources. It is also the point where the computational burden of a standard Gaussian process becomes non-trivial, and many practitioners will start to consider faster, more scalable alternatives. 

The $(n=1000, NSR=0)$ and $(n=1000, NSR=0.1)$ cases are summarized, respectively, by \cref{fig:rank_1a,fig:rank_1b} using cumulative rank plots across the 600 simulation scenarios (60 test functions repeated 10 times each). Curves that are concave downward represent methods that frequently outperform others and rarely rank near the bottom (e.g., \emu{rgasp} in \cref{fig:rank_1a}), while concave up (e.g., \emu{baseline}) usually rank near the bottom. An ``S"-shaped curve suggests a robust method that rarely ranks at the top or bottom (e.g., \emu{mpgp}), while a reverse ``S" shape (e.g., \emu{treegp} in \cref{fig:rank_1a}) indicates a method with highly variable performance across scenarios. Legend order and color assignments reflect the area under each curve, equivalent to average rank.

Gaussian process emulators tend to perform quite well, with \emu{rgasp} placing as the top ranked emulator for a large number of simulation scenarios. Among the ``fast" emulators, \emu{svecgp} and \emu{bppr} appear to be competitive across the various cases. 

\Cref{fig:heat_1a,fig:heat_1b} extend this analysis by presenting results function by function, using numerical median CRPS values (averaged across replications) rather than ranks. The emulator labels on the x-axis of these plots are ordered from left to right by the median CRPS averaged across all test functions. While the rank plots highlight how often one emulator outperforms another, the heatmaps provide insight into the magnitude of those differences broken out by test function.

To more explicitly examine the trade-off between predictive accuracy and computational cost, \cref{fig:combined} shows a Pareto front plot of average relative CRPS versus runtime. Relative CRPS is defined as described earlier, with a small offset to reduce sensitivity to negligible differences and a cap to limit the influence of outliers. Emulators near the lower-left corner of the plot represent favorable accuracy–efficiency tradeoffs, as no other method is both faster and more accurate on average across scenarios.

Several emulators not yet discussed become interesting under the lens of this tradeoff. For example, the \emu{rffgp}, \emu{confrf}, and even \emu{blm} emulators may be worth a closer look for some problems, when computational efficiency is paramount. We also observe this trade-off within the standard GP implementations: \emu{rgasp} tends to be more accurate than \emu{gp}, but it is generally slower. 

Together, these figures illustrate both how often one emulator outperforms another and by how much. The rank plots capture frequency of dominance, but not magnitude; the heatmaps and Pareto fronts help fill in that missing context. Additional figures using variations on these metrics are given in the supplement. For deeper scenario-level analysis, including boxplots and case studies, see \cref{sec:discuss}.

\begin{figure}[h]
\centering
\includegraphics[
  alt={Cumulative rank curves showing several competitive emulators with 500 training points and no single dominant method. See caption for details.},
  width=0.98\linewidth
]{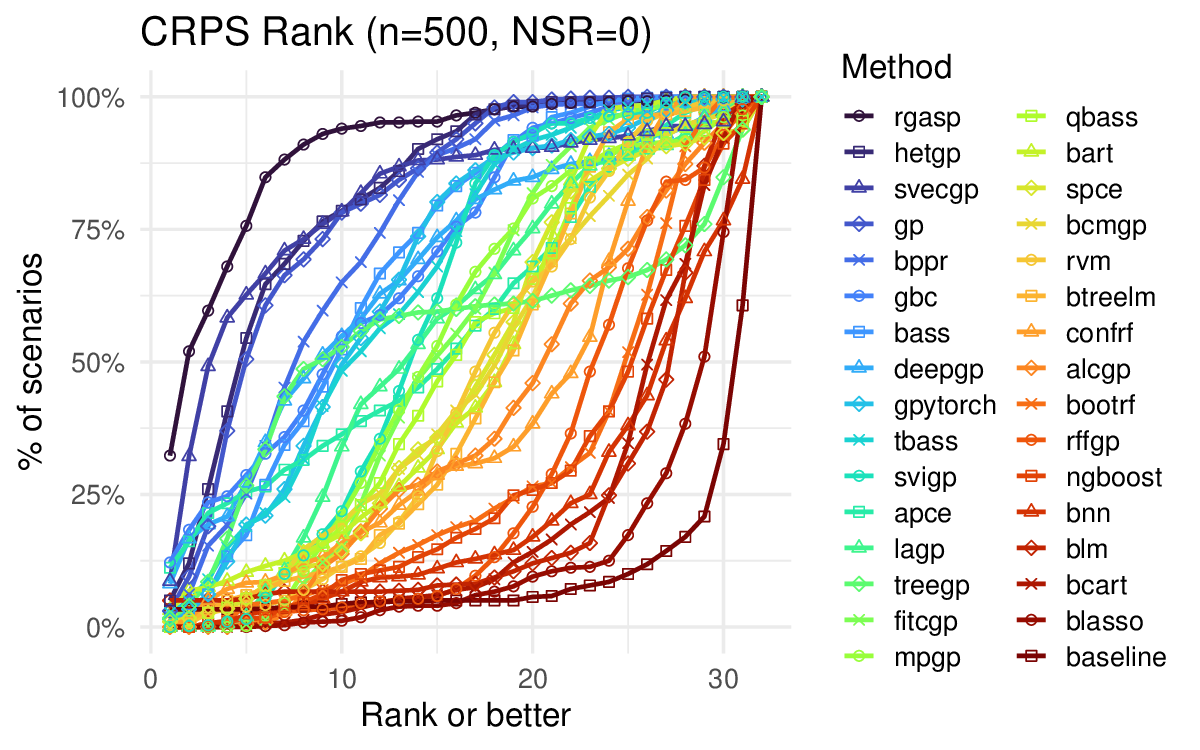}
\caption{Cumulative rank-plot for $n=500$, $NSR=0$ setting. The curve for each emulator shows the proportion of cases that the method was at least top $r$ out of $32$ in terms of CRPS, for $r=1,\ldots, 32$. }
\label{fig:rank_2}
\end{figure}

\subsection{Small Training Set (n = 500)}

The case where only $n = 500$ simulator evaluations are available is common when dealing with expensive models or limited computational budgets. In \cref{fig:rank_2}, we show cumulative rank plots summarizing emulator performance in this regime, restricted to the noise-free setting ($\text{NSR} = 0$). A full set of results, including heatmaps, timing comparisons, and the $\text{NSR} = 0.1$ case, can be found in the supplemental materials.

\begin{figure}[h]
\centering
\includegraphics[alt={ Cumulative rank curves showing several competitive scalable emulators with 5000 training points. See caption for details. }, width=0.98\linewidth]{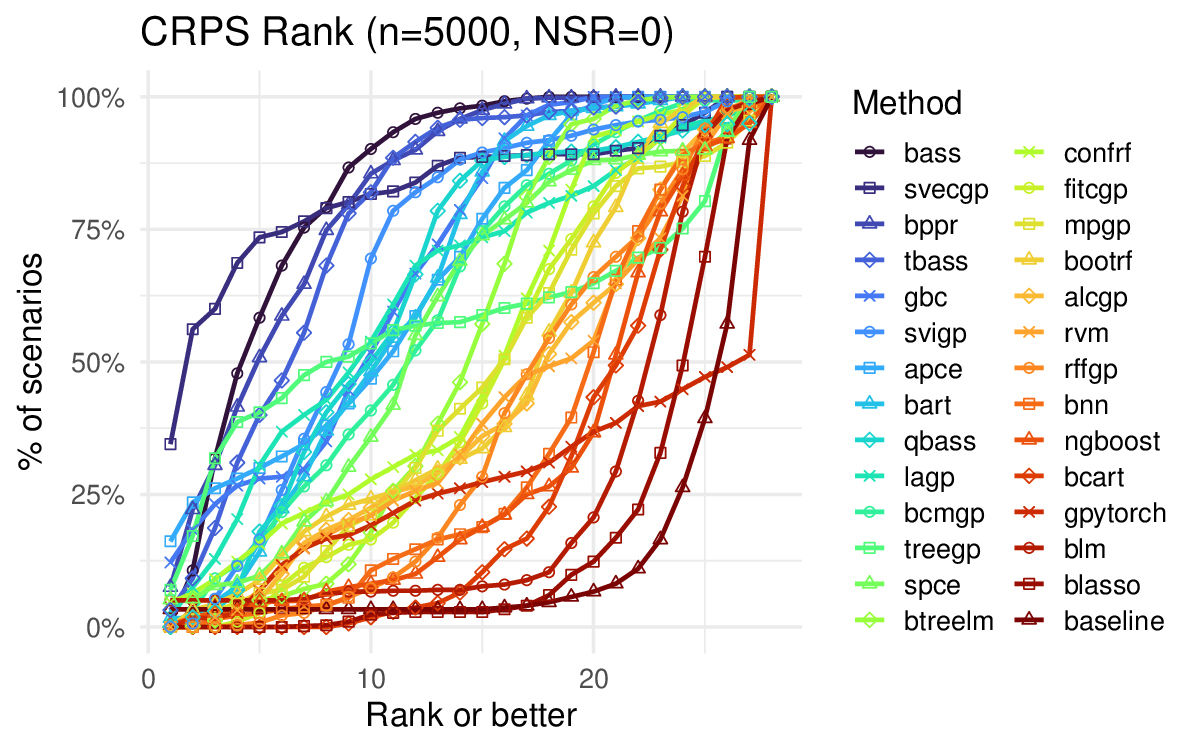}
\caption{Cumulative rank-plot for $n=5000$, $NSR=0$ setting. The curve for each emulator shows the proportion of cases that the method was at least top $r$ out of $32$ in terms of CRPS, for $r=1,\ldots, 32$. }
\label{fig:rank_3}
\end{figure}

\subsection{Large Training Set (n = 5000)}

When $n = 5000$ evaluations are available, the design space is typically better-explored and computational constraints become more prominent. \Cref{fig:rank_3} provides a high-level overview of emulator rankings in the noise-free setting. Due to runtime costs and practical relevance, we omit the noisy case here; additional figures for this regime are available in the supplement.

\begin{figure}[h]
\centering
\includegraphics[alt={ Performance profiles showing substantial differences in emulator accuracy and consistency across small datasets. See caption for details. }, width=0.98\linewidth]{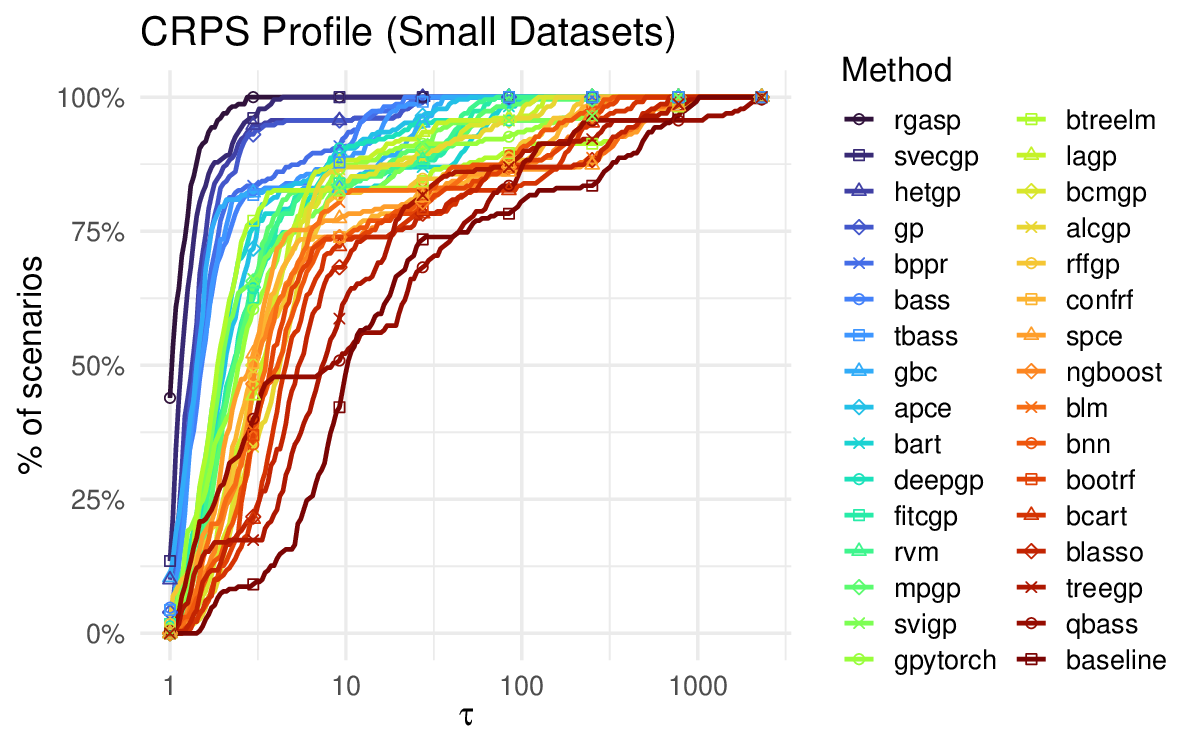}
\caption{Performance profile plots for the real datasets with training size $n < 2000$. For each emulator, the curve shows the proportion of datasets for which the method is within a factor $\tau$ of the best-performing emulator in terms of CRPS. Curves closer to the upper-left corner indicate stronger and more consistent performance. Note that \emu{treegp} only completed on eleven datasets in the allotted time and was replaced by the baseline model for the remaining cases.}
\label{fig:rank_4}
\end{figure}

\begin{figure}[h]
\centering
\includegraphics[alt={ Performance profiles showing substantial differences in emulator accuracy and consistency across large datasets. See caption for details. }, width=0.98\linewidth]{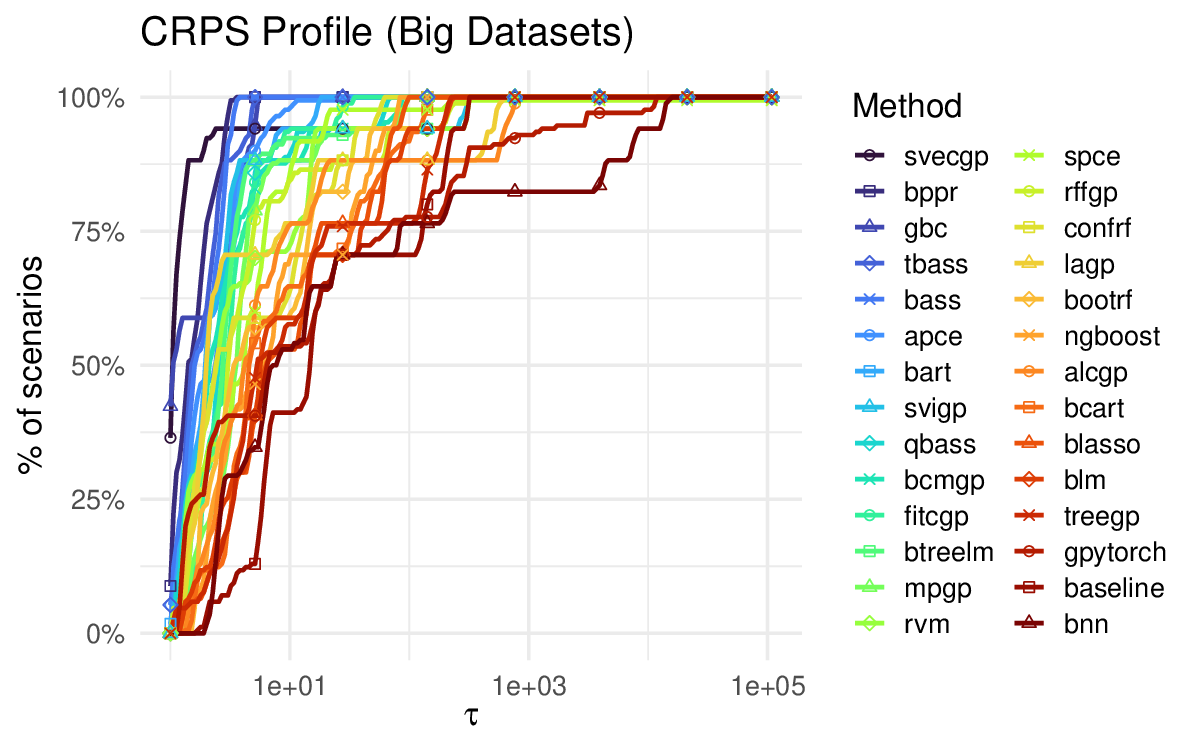}
\caption{Performance profile plots for the real datasets with training size $n \geq 2000$. For each emulator, the curve shows the proportion of datasets for which the method is within a factor $\tau$ of the best-performing emulator in terms of CRPS. Curves closer to the upper-left corner indicate stronger and more consistent performance. Several slower emulators were excluded from this analysis.}
\label{fig:rank_5}
\end{figure}

\subsection{Real Data}

For the real datasets included in our analysis, \cref{fig:rank_4} and \cref{fig:rank_5} show performance profiles rather than rankplots (see discussion in \cref{sec:results} and SM5.1.2 of the supplement for details). The rank-based summaries can be viewed in Section SM5.2.1 of the supplement. We remind the reader that these results reflect the implementation and hyperparameter choices for each emulator as much as the underlying approach. The tradeoff between accuracy and computation can often be explicitly controlled (e.g., the size of the conditioning sets for \emu{svecgp} or the number of training iterations for \emu{gpytorch}). We therefore emphasize again that the comparisons here reflect these choices, rather than fully tuned performance. 

\begin{figure}[ht]
  \centering
  \begin{subfigure}[t]{0.48\textwidth}
    \centering
    \includegraphics[alt={ Pareto plots showing different accuracy and computational-cost tradeoffs for small datasets. See caption for details. }, width=\linewidth]{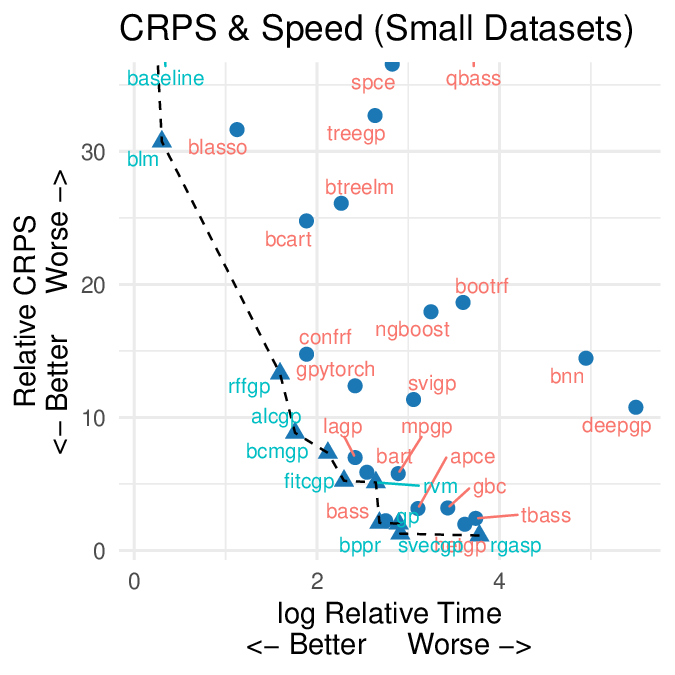}
    \caption{}
    \label{fig:paretosmall}
  \end{subfigure}
  \hfill
  \begin{subfigure}[t]{0.48\textwidth}
    \centering
    \includegraphics[alt={Pareto plots showing different accuracy and computational-cost tradeoffs for large datasets. See caption for details.  }, width=\linewidth]{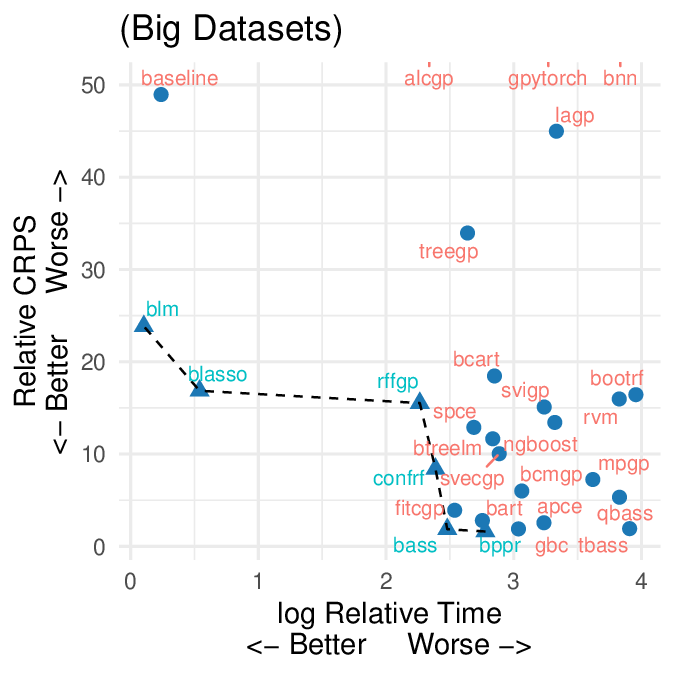}
    \caption{}
    \label{fig:paretobigg}
  \end{subfigure}
  \caption{Pareto front of emulators based on average relative CRPS and log relative runtime across all simulation scenarios. Relative CRPS is computed as ($\text{CRPS} + \omega) / (\min \text{CRPS} + \omega$) for each scenario, capped at $1000$ to reduce sensitivity to outliers ($\omega = 0.001$). The dashed line and triangle markers indicate the methods which are not dominated (slower and less accurate) by any other emulator.}
  \label{fig:paretodata}
\end{figure}

For these computer models, the \emu{svecgp} stands out as a highly effective emulator. Across all datasets (comparing to only the fast emulators), \emu{svecgp} achieves the smallest mean CRPS in $45.0\%$ of all folds. Had we ranked by median CRPS or the max-CRPS (across the test set), the win rate drops to $34.0\%$ and $36.0\%$, respectively; still indicating strong performance, but less conclusively. And, of course, these figures do not tell us how {\it much} better it is in those cases, or whether the difference is practically meaningful.  For instance, the relative CRPS in \cref{fig:paretobigg} paints \emu{svecgp} as somewhat less effective (in the big data setting), likely driven by its poor performance on a few datasets, like the high-dimensional dataset \tf{nuclear\_data}. On the other hand, an emulator like \emu{bass}, which demonstrates the best performance just $3.25\%$ of the time, seems to perform reasonably well across all datasets, ranking in the top half of emulators $97.8\%$ of the time, more than any other method. The relative mean CRPS (with $\omega = 0.001$; see \cref{eq:rel_crps}) for \emu{bass} is $1.8$, which is substantially better than even \emu{svecgp} with a relative mean CRPS of $4.9$. At a minimum, this suggests that \emu{bass} is a robust and effective algorithm at its default settings; it is quite likely that with sophisticated tuning, many of the other methods could demonstrate similar proficiency across the board. 

\begin{figure}[h]
    \centering
    \includegraphics[alt={ Heatmap showing substantial variation in emulator performance across small real-world datasets, with no uniform winner. See caption for details. }, width=0.98\linewidth]{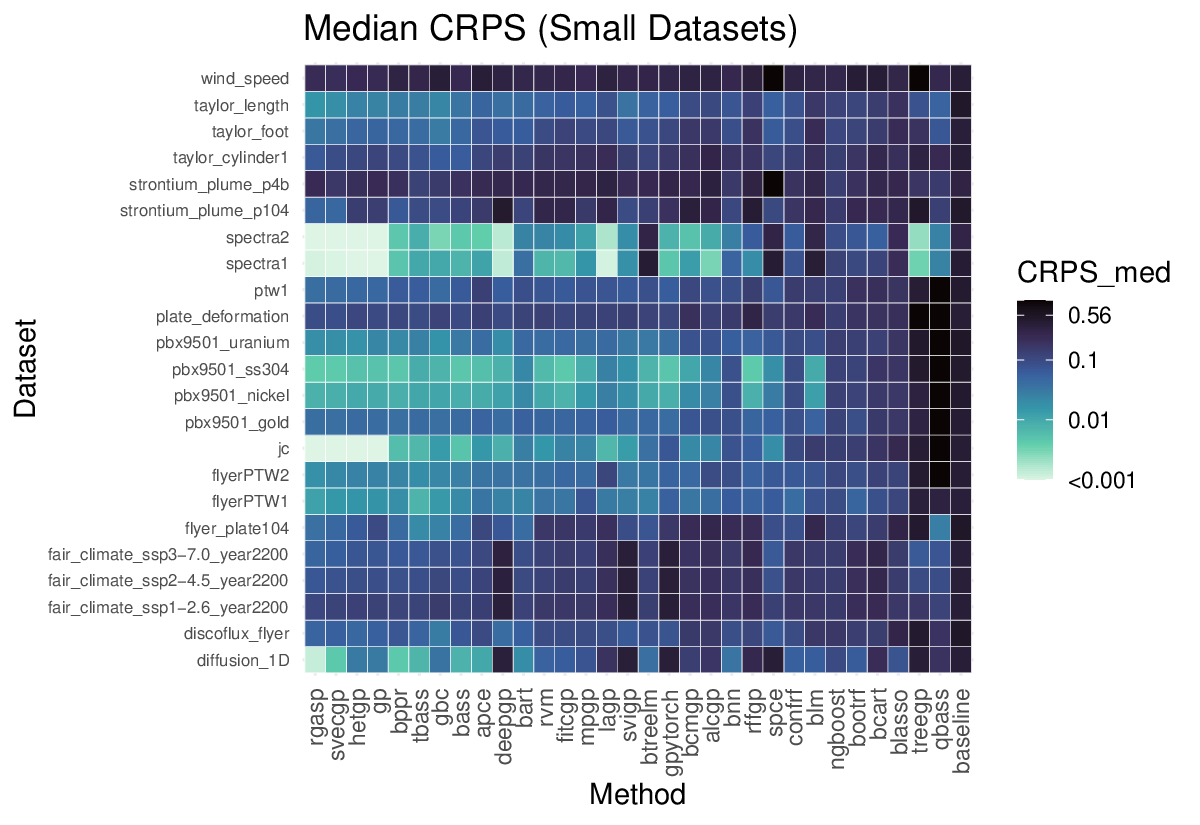}
    \caption{A heatmap of CRPS (truncated between $0.001$ and $1.0$ for visual simplicity) for the real datasets with $n < 2000$.}
    \label{fig:heat_1c}
\end{figure}

\begin{figure}[h]
    \centering
    \includegraphics[alt={ Heatmap showing substantial variation in emulator performance across large real-world datasets, with no uniform winner. See caption for details. }, width=0.98\linewidth]{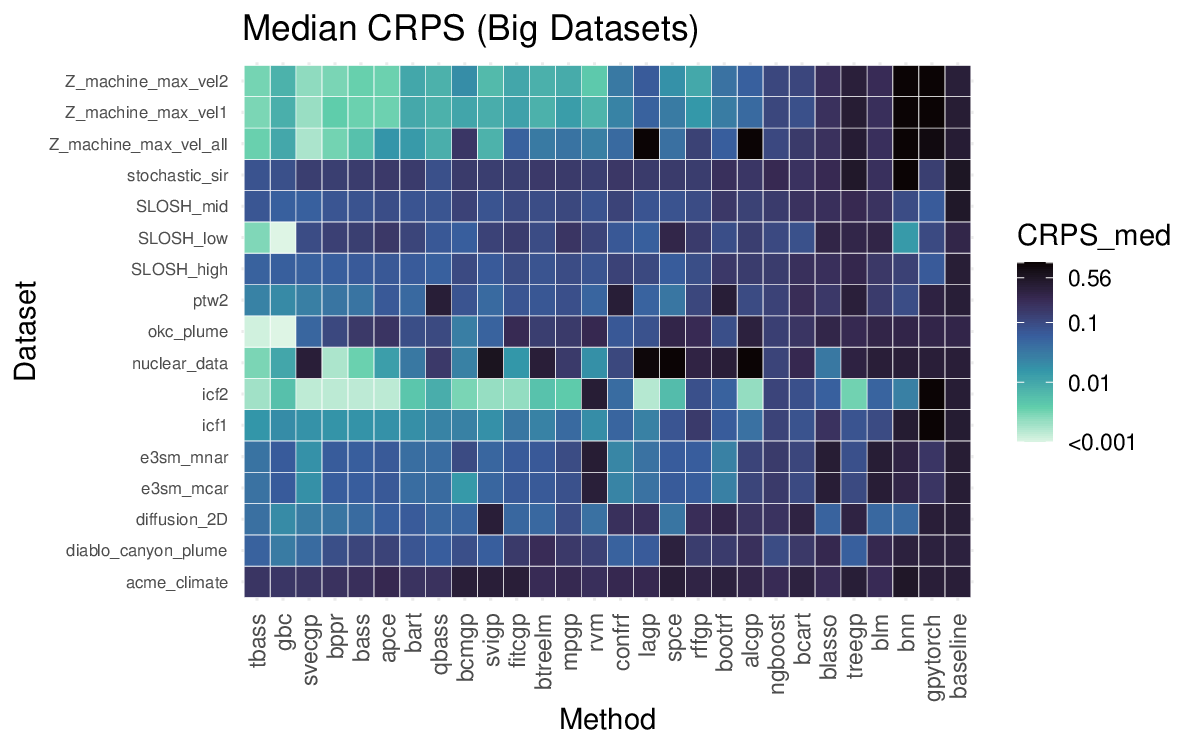}
    \caption{A heatmap of CRPS (truncated between $0.001$ and $1.0$ for visual simplicity) for the real datasets with $n \geq 2000$.}
    \label{fig:heat_1d}
\end{figure}

\section{Discussion and Analysis}
\label{sec:discuss}

\begin{figure}[ht]
  \centering
  \begin{subfigure}[t]{0.48\textwidth}
    \centering
    \includegraphics[alt={CRPS boxplots for the ignition test function, showing deep GP as the strongest performer. See caption for details.}, width=\linewidth]{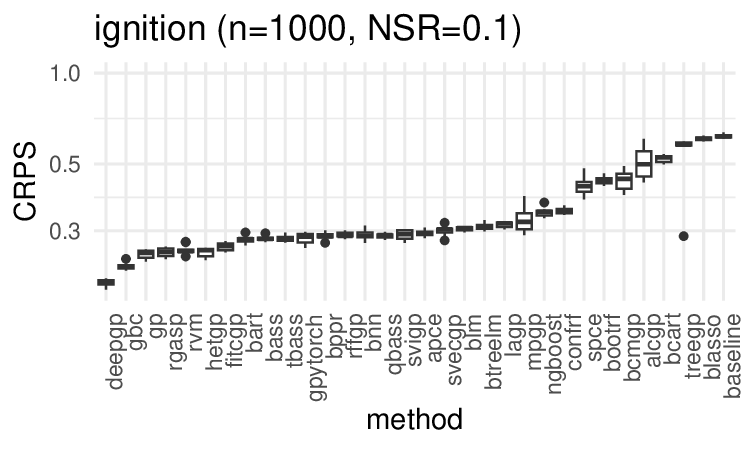}
    \caption{}
    \label{fig:box_ignition}
  \end{subfigure}
  \hfill
  \begin{subfigure}[t]{0.48\textwidth}
    \centering
    \includegraphics[alt={CRPS boxplots for the banana test function, showing adaptive polynomial chaos as the strongest performer. See caption for details.}, width=\linewidth]{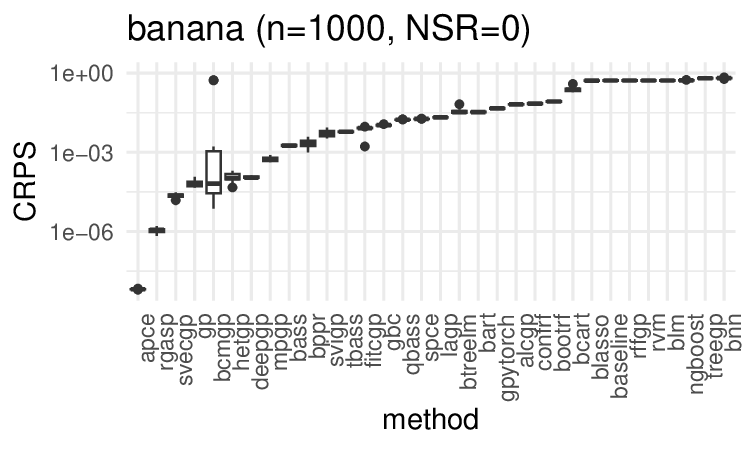}
    \caption{}
    \label{fig:box_banana}
  \end{subfigure} 

\vspace{0.5em}
  
  \begin{subfigure}[t]{0.48\textwidth}
    \centering
    \includegraphics[alt={CRPS boxplots for the cube5 test function, showing BART as the strongest performer. See caption for details.}, width=\linewidth]{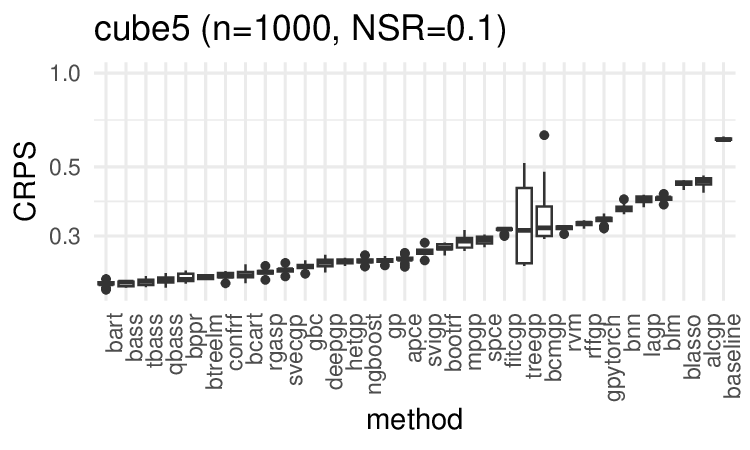}
    \caption{}
    \label{fig:box_friedman}
  \end{subfigure}
  \hfill
  \begin{subfigure}[t]{0.48\textwidth}
    \centering
    \includegraphics[alt={CRPS boxplots for the multivalley test function, showing BASS as the strongest performer. See caption for details.}, width=\linewidth]{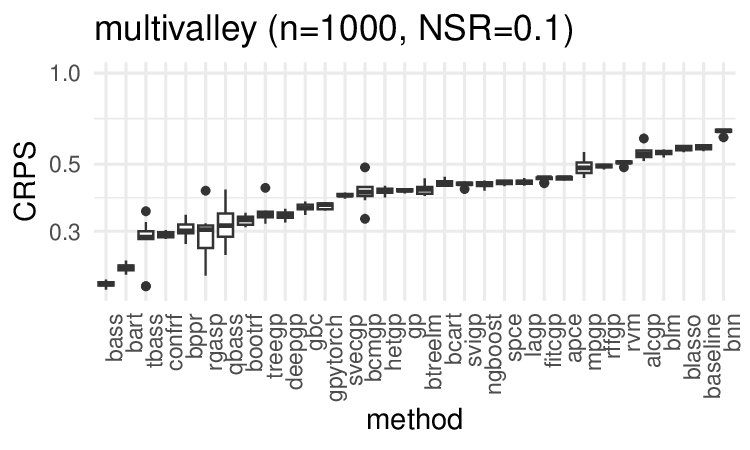}
    \caption{}
    \label{fig:box_cube}
  \end{subfigure} 

  \vspace{0.5em}

\begin{subfigure}[t]{0.48\textwidth}
    \centering
    \includegraphics[alt={CRPS boxplots for the Z-machine maximum-velocity dataset, showing scaled Vecchia GP as the strongest performer. See caption for details.}, width=\linewidth]{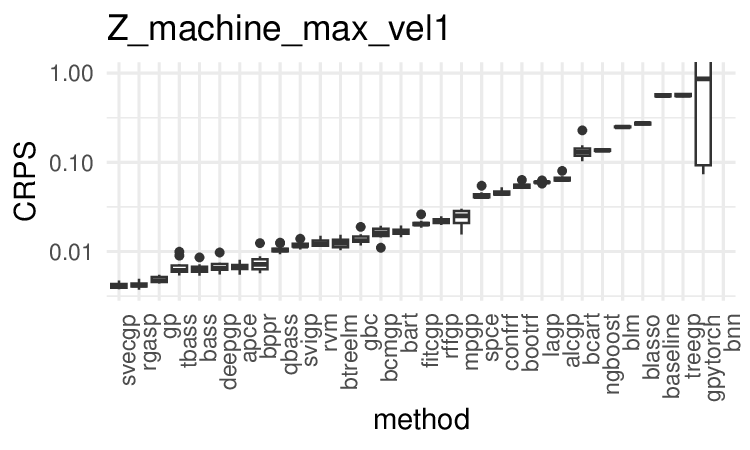}
    \caption{}
    \label{fig:box_z}
  \end{subfigure}
  \hfill
  \begin{subfigure}[t]{0.48\textwidth}
    \centering
    \includegraphics[alt={CRPS boxplots for the nuclear-data dataset, showing Bayesian projection pursuit regression as the strongest performer. See caption for details.}, width=\linewidth]{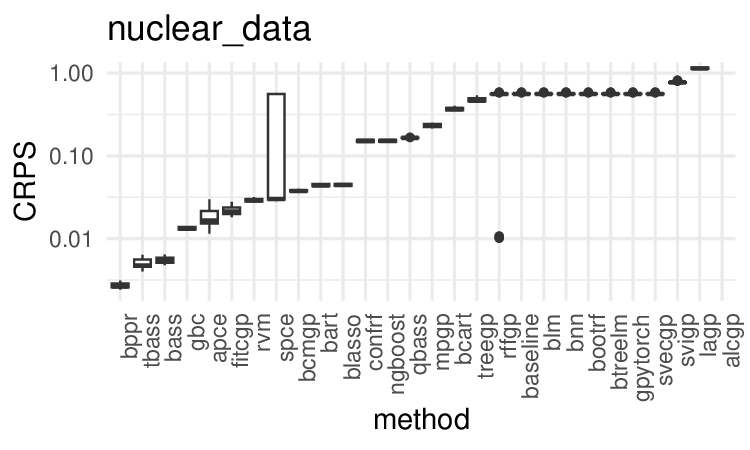}
    \caption{}
    \label{fig:box_n}
  \end{subfigure} 
  \caption{Boxplots of CRPS across ten replications/folds for four test functions and two real datasets. In each panel, a different emulator seems to be the ``best" for that particular scenario.} 
  \label{fig:boxplots}
\end{figure}

\subsection{There is no free lunch (or best emulator)}

In uncertainty quantification, it is common to ask: {\it which emulator is the best?} But this is rarely the most productive question to pursue. The no free lunch theorems \citep{wolpert2002supervised} guarantee that no emulator can possibly ``win" on all functions. Indeed, of the $32$ emulators in this study, $28$ of them had the lowest CRPS value on at least one of the simulation scenarios. Six (out of many) notable cases are shown in \cref{fig:boxplots}, demonstrating that different emulators can dominate for different functions. The \emu{deepgp} emulator is an interesting case study: this emulator does reasonably well in the synthetic data simulation study with an overall win rate of $9.6\%$ and an average ranking of $11.0/32$. But \emu{deepgp} was not designed to be a catch-all emulator; it was designed to be effective for challenging nonstationary emulation problems where sophisticated tools are required and, in these settings, it does exceptionally well. On test functions such as \tf{foursquare}, \tf{squiggle}, \tf{star2}, and \tf{ignition}, it often exhibits the best CRPS by a wide margin (see \cref{fig:box_ignition} for an example).

For another example, consider two of the standout fast emulators in this study: \emu{svecgp} and \emu{bppr}. The \emu{svecgp} emulator had an overall win rate of $11.8\%$ in the synthetic data simulations, with an average rank of $7.2$ out of $32$. By comparison, \emu{bppr} achieved the best CRPS in just $3.9\%$ of scenarios and had an average rank of $7.7$. However, the average CRPS relative to the best model (see \cref{eq:rel_crps}; $\omega=0.001$) was $32.9$ for \emu{svecgp} and only $10.0$ for \emu{bppr}, suggesting that \emu{bppr} was often more consistently and practically competitive even when it did not win outright. Stratifying by noise-to-signal ratio reveals a modest shift in their relative behavior: \emu{svecgp} has a head-to-head win rate of $70.0\%$ against \emu{bppr} when $NSR=0$, but only $48.7\%$ when $NSR=0.1$. The corresponding average ranks tell a similar story: \emu{svecgp} averaged $7.0$ and $7.6$ in the $NSR=0$ and $NSR=0.1$ settings, compared with $8.1$ and $7.1$ for \emu{bppr}. Taken together, these results reinforce the no-free-lunch concept. Even among strong fast emulators, relative performance depends on the structure of the problem, with \emu{svecgp} favored more often in noise-free settings and \emu{bppr} becoming more competitive when noise is present.

These results suggest that searching for a single emulator that performs best across all functions is a fruitless endeavor. Moreover, searching for the best emulator in a {\it single} setting may also be misguided. ``{\it Amazing things come from having many good models}" and ``{\it all models are wrong but {\bf many} are useful}" describe the sentiment behind the Rashomon effect, an idea that has gained increasing attention in recent years \citep{fisher2019all, xin2022exploring, rudin2024amazing, visokay2025you}. The term `Rashomon effect' was originally coined by Leo Breiman to describe the phenomenon in which multiple distinct models explain the data equally well, yet may lead to different conclusions \citep{breiman2001random}. When this occurs (in sensitivity analysis or Bayesian model calibration, for example) it may signal underlying limitations in the modeling assumptions or data, including potential issues related to parameter identifiability or model discrepancy. Conversely, agreement among similarly accurate (yet distinct) emulators can increase confidence in the results. A short case study illustrating this idea is provided in the supplement.

\subsection{Tuning an emulator}
\label{sec:tune}
\begin{figure}[h]
    \centering
    \includegraphics[alt={ Pareto plot showing that nugget choice substantially affects local Gaussian process performance and runtime. See caption for details. }, width=0.98\linewidth]{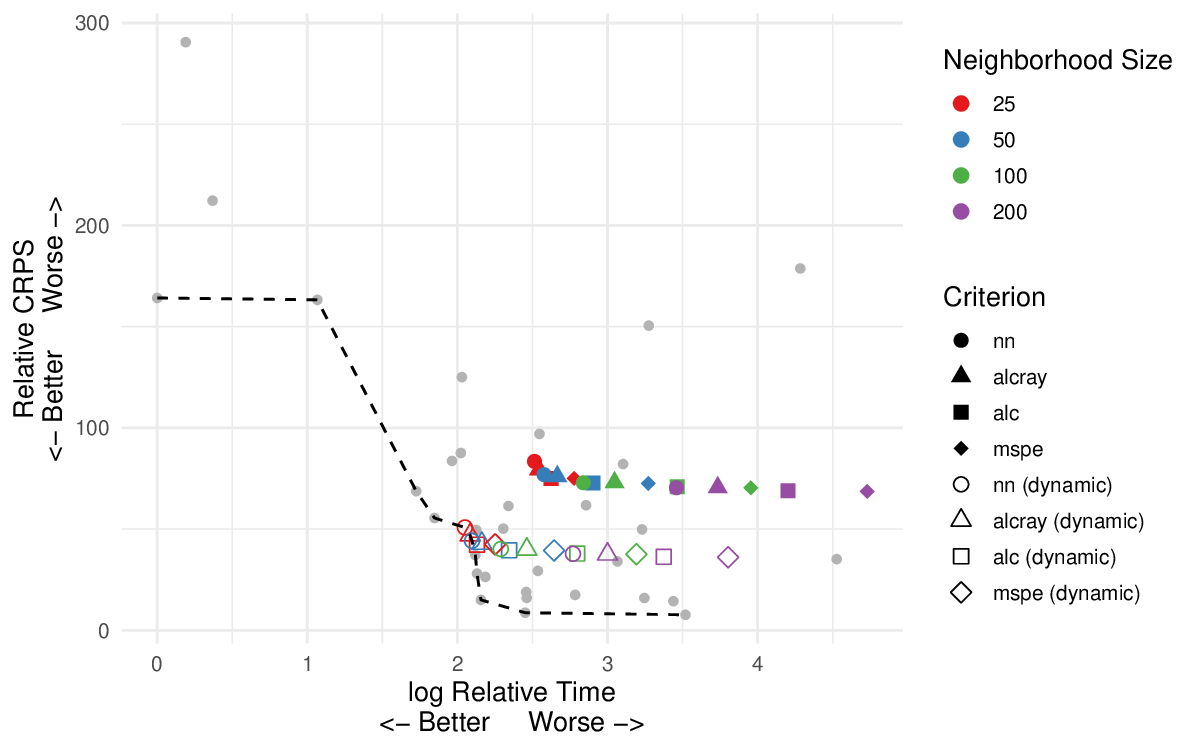}
    \caption{The Pareto front of \cref{fig:pareto1} is recreated here, with the addition of $32$ new \emu{lagp} emulators with various hyperparameter values.}
    \label{fig:pareto_lagp}
\end{figure}

To demonstrate how \texttt{duqling} can support exploration of tuning parameters, we re-ran the simulation study with 32 variants of the \emu{lagp} emulator, modifying three key parameters: the neighborhood size, the local design criterion, and the nugget size. Specifically, we considered neighborhood sizes of $25$, $50$, $100$, and $200$, and four selection criteria---\texttt{nn}, \texttt{alcray}, \texttt{alc}, and \texttt{mspe}---ranging from fastest to slowest in terms of computational cost (see \cite{gramacy2015local} for details). We also compared two different nugget values: the default value of $0.0001$ and a data-scaled nugget value (usually smaller, as suggested by the \emu{lagp} documentation) equal to the response variance divided by $10^7$. These variants were treated as new emulators and {\it joined} with the existing simulation results (see \cref{sec:duqling}). 

\Cref{fig:pareto_lagp} shows the resulting Pareto front. The dynamic nugget choice does lead to a substantial improvement in performance (and computational efficiency) in this setting. Interestingly, while neighborhood size has a noticeable impact on runtime, its effect on CRPS performance is relatively small. The same is broadly true for the selection criterion, though its impact on performance appears more pronounced when the neighborhood size is small. For this metric and set of test functions, a favorable tradeoff between speed and accuracy is achieved by choosing either (i) the more sophisticated \texttt{alc} criterion with a neighborhood size of 25, or (ii) the faster \texttt{nn} criterion with a neighborhood size of 100. 

It is worth noting that the results in this section are somewhat sensitive to the choice of metric. For example, if the relative CRPS value is calculated with $\omega=0$ (see \cref{eq:rel_crps}), then the differences between methods become negligible, and the default nugget is even preferred in some cases. See \url{github.com/knrumsey/duqling_results/R/duqling/tuning.R} for details and for additional exploration of alternative metrics.

While this example focuses on \emu{lagp}, the same approach can be applied to any emulator. The ability to treat tuned variants as distinct methods makes it easy to identify effective default settings, or to study how emulator behavior responds to different hyperparameters. This can be a valuable tool both for practitioners selecting models and for researchers developing new emulation strategies.

\begin{figure}[ht]
  \centering
  \begin{subfigure}[t]{0.48\textwidth}
    \centering
    \includegraphics[alt={ Multidimensional scaling plot showing distinct groups of emulators with similar performance patterns. See caption for details. }, width=\linewidth]{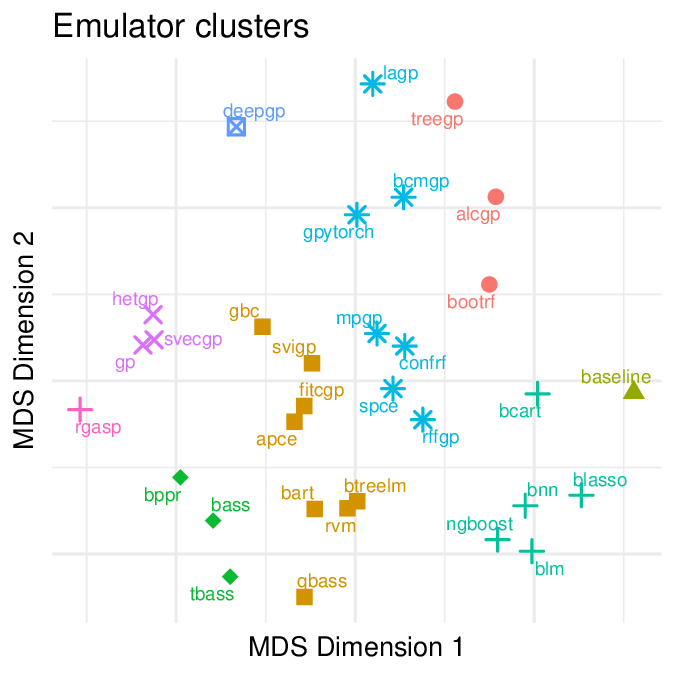}
    \caption{A clustering of emulators using rank-based performance similarity and multidimensional scaling. Nearby emulators performed similarly on the various datasets and test functions.}
    \label{fig:clust1}
  \end{subfigure}
  \hfill
  \begin{subfigure}[t]{0.48\textwidth}
    \centering
    \includegraphics[alt={ Multidimensional scaling plot showing groups of datasets that elicit similar emulator performance patterns. See caption for details. }, width=\linewidth]{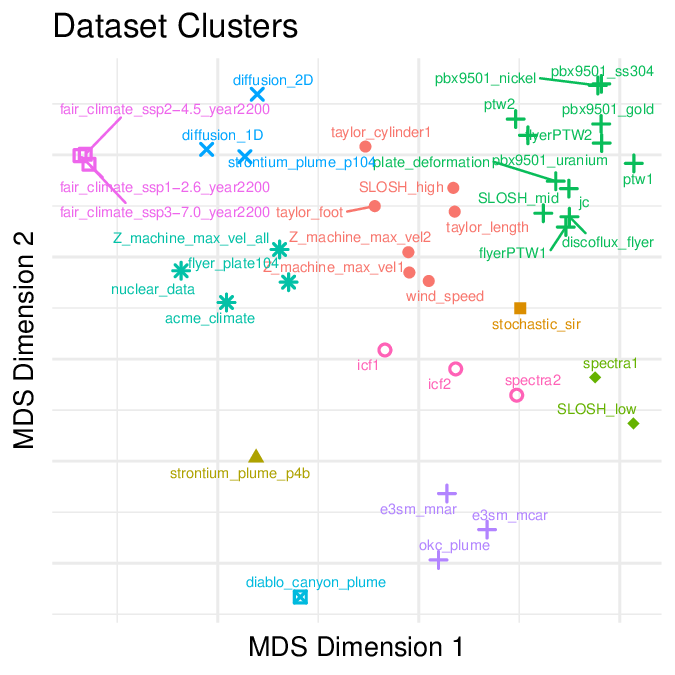}
    \caption{A clustering of datasets using rank-based performance similarity and multidimensional scaling. Nearby datasets led to similar performance by the various emulators.}
    \label{fig:clust2}
  \end{subfigure}
  \caption{}
  \label{fig:clusters}
\end{figure}

\subsection{Clustering by performance}
\label{sec:cluster}
We also explore the structure of these results through a simple clustering analysis. Rather than focusing on individual comparisons, we ask {\it which emulators behave similarly across scenarios, and which datasets elicit similar behavior from the emulators?} To accomplish this, we construct rank-based performance vectors for emulators (across all datasets and test functions) and for datasets (across all emulators), then compute pairwise distances between these vectors. We visualize the resulting geometry using classical multidimensional scaling and identify clusters using DBSCAN. \Cref{fig:clusters} illustrates the results of this analysis using multidimensional scaling and DBSCAN clustering \citep{ester1996density}. 

On the emulator side, distinct classes of methods emerge, and even within those families the relative distances carry information. The basis function approaches \emu{bppr} and \emu{bass} sit tightly together, with related methods such as \emu{rffgp}, \emu{apce}, and \emu{bart} somewhat nearby. Tree-based methods form another distinct group in the bottom right. Several Gaussian process clusters appear: the SoD-adjacent methods \emu{alcgp}, \emu{mpgp}, \emu{fitcgp}, and \emu{bcmgp} are all close together; \emu{deepgp} and \emu{treegp} show up as nearest neighbors; and the classical \emu{gp} and its heteroskedastic counterpart \emu{hetgp} naturally cluster. For a new dataset, \cref{fig:clust1} may help a practitioner identify a sensible set of candidate emulators. While considering $32+$ candidates may be unnecessary, it is often reasonable to include at least one representative from each major group.

On the dataset side, clustering reveals natural classes of problems and also points toward opportunities for future work. For any given emulator, it is useful to know which classes of datasets it tends to succeed on, and which expose its weaknesses. In some cases it may make sense to double down on problem types where a method already excels, while in others one might identify weak spots and build targeted extensions. For example, if one cluster of datasets is consistently challenging for an entire family of emulators, that cluster becomes a clear target for specialized method development (the \tf{strontium\_plume\_p4b} dataset seems to be a ripe challenge). Another potential use is to match synthetic test functions to real datasets: functions that cluster closely with observed data may be especially valuable as benchmarks. For brevity, additional clustering plots, including one that combines both test functions and datasets, are provided in the supplement.

\subsection{How to pick an emulator}
If the practitioner takes just one thing away from this work, it should be that there are many choices and considerations when it comes to choosing an emulator. We strongly encourage examining a suite of candidates, at least $1$ from each cluster in \cref{fig:clust1}, before choosing one (or many, as suggested by the Rashomon effect). 

For many users, predictive accuracy is the primary (if not the only) criterion for selecting an emulator. Some emulators perform exceptionally well on specific functions and possibly not so well on others. Some emulators perform well across a broad range of scenarios despite rarely being ``the best". Depending on how well one understands the underlying data, this tradeoff between robustness and peak performance may be an important consideration.

Of course, there are many reasons to choose an emulator that have less to do with raw accuracy. When the simple \emu{blm} works well enough, it brings with it at least a century of statistical insight. Everything from uncertainty quantification to hypothesis testing to experimental design has been deeply studied in that framework. Similarly, polynomial chaos expansions make global sensitivity analysis via Sobol indices essentially trivial \citep{sudret2017surrogate}, though other emulators like \emu{bass} and \emu{bart} can also provide closed-form estimates \citep{francom2018sensitivity, zamanian2021significant, horiguchi2023estimating}.

Some emulators are better suited to particular modeling goals. For example, Gaussian processes and basis regression models support closed-form active subspace directions \citep{constantine2015activebook}, which can help uncover low-dimensional structure \citep{wycoff2021sequential, rumsey2024discovering}. Other methods are designed around specific UQ tasks—such as expected improvement for Bayesian optimization \citep{jones1998efficient}, surrogate modeling across multiple simulators \citep{yannotty2024model, motamed2020multi}, simulators with categorical inputs \citep{zhang2021mixed, francom2019inferring}, or analyzing adjacent computer models \citep{bernstein2019comparison, rumsey2025coactive}. A related trend is the development of physics-informed surrogates, which can outperform purely data-driven models by incorporating structural knowledge \citep{mcclarren2011physics, kashinath2021physics, dalton2023physics, orjuela2026interpretable}. Sometimes specialized emulators are required to accommodate particular data structures \citep{da2012gaussian, mallasto2018wrapped, flowers2026modular}. In other cases, emulators are popular because they are familiar and trusted by a community: \emu{rvm} extends support vector machines, and \emu{confrf} builds on random forests using conformal inference for fast model-free UQ. Stochastic simulators often require specialized treatment, and much research has focused on this challenge \citep{baker2020predicting, rumsey2024generalized, binois2021hetgp, plumlee2014building, oakley2017calibration, zhu2023stochastic, luthen2023spectral}. Functional or vector-valued outputs (e.g., spatial or temporal fields) present additional challenges for emulation, and often require specialized modeling strategies or extensions of standard surrogate methods \citep{liu2004dynamic, higdon2008computer, conti2010bayesian, gu2016parallel, francom2025elastic}. Many important applications fall into this class, and extending the framework to these settings is an important direction for future work in \texttt{duqling}. Emulation with high-dimensional inputs (e.g., $p>n$) may also require customization, though approaches that perform a model search like tree-based or basis function models can work well without modification \citep{chakraborty2017emulation,lataniotis2020extending, rumsey2024discovering, shustin2026pce}. Hybrid approaches, which combine elements of various emulator types, are also a common approach for exploiting strengths and mitigating weaknesses. For example, Gaussian processes have been combined with spline representations \citep{chakraborty2013spline,bowman2016emulation}, polynomial chaos expansions \citep{schobi2015polynomial,zhou2019new}, and Bayesian regression trees \citep{maia2024gp}.

Ultimately, the right emulator may depend critically on the task it is intended for. If training is a one-time cost and downstream inference is fast, a more complex, state-of-the-art model may be justified \citep{radev2023jana}. If hundreds of emulators are required, a training-free method like \emu{lagp} can be especially appealing \citep{gramacy2015local}. In settings requiring millions of sequential predictions, the choice of emulator should reflect that constraint \citep{rumsey2023localized, braun2024gptreeo, hutchings2024fast}. In this sense, downstream context should shape your emulator selection just as much (if not more) than eking out that last percent of CRPS performance.

\section{Conclusion}
\label{sec:conclusions}

This paper presents a large-scale, reproducible comparison of 32 emulators across 60 synthetic test functions and 40 real-world datasets. While no single emulator dominates in all settings, the results highlight important tradeoffs between accuracy, robustness, and runtime. Notably, \emu{gp} \citep{binois2021hetgp} and \emu{rgasp} \citep{gu2018robust} both demonstrate strong predictive performance, with some evidence of a time–accuracy tradeoff between them. The \emu{svecgp} emulator stands out for its excellent scaling and reliable accuracy across a wide range of scenarios \citep{katzfuss2022scaled}. We also find that \emu{bppr} performs consistently well across a broad range of settings, with particularly strong performance in the presence of noise \citep{collins2024bayesian}.

The \texttt{duqling} framework, available at \url{github.com/knrumsey/duqling} and \url{github.com/reidmorris/duqling_py}, played a central role in enabling this study and is designed to support future emulator benchmarking, tuning, and development. All results in this paper were generated using open infrastructure that can be extended or modified to suit new problems, helping to make emulator evaluation both rigorous and reproducible.

A natural next step is to extend this framework to a broader class of emulation problems. While this study focuses on scalar-output, deterministic simulators with continuous inputs, many important applications involve additional complexities. In particular, future work will involve extending this framework to conduct automated, large-scale, and reproducible comparisons in settings with categorical inputs, stochastic simulators, and multivariate or functional outputs. These settings introduce new modeling challenges and tradeoffs, and we expect that some of the lessons from this study will carry over, while also highlighting differences in emulator behavior in these more complex settings.

We acknowledge that several of the emulators evaluated in this paper, including \emu{bass}, \emu{bppr}, and \emu{apce}, were developed in part by the authors of this study. We have tried to be transparent and fair in our analysis, with the goal of supporting reproducible comparisons. Our motivation is to move the field forward, not to push our own methods, and we are excited by the success (and the potential for even greater future success) of emulators developed by others in the UQ community. We encourage the interested reader to compare their own results to those in this paper using \texttt{duqling}.

\section*{Supplemental Materials}

Additional details, figures, and implementation information are provided in the supplement. Section SM1 contains the R scripts used to reproduce all results. Section SM2 summarizes emulator performance across all simulation scenarios, including overall summaries and breakdowns by test function and dataset. Section SM3 provides technical details on the reproducible simulation study framework, including seed construction, fallback models, output formats, and descriptions of the datasets and test functions. Section SM4 presents additional discussion of the Rashomon effect and ensembling. Section SM5 includes extended figures and analyses for multiple performance metrics (e.g., FVU, CRPS, and runtime), along with additional diagnostics such as performance profiles, heatmaps, and boxplots.

\section*{Acknowledgements}
We would like to thank the editors, associate editor, and the two anonymous reviewers for their feedback, which has led to a much improved paper. We are also grateful to Derek Bingham, Gavin Collins, Annie Booth, Matthias Katzfuss, Bruno Sudret, Bobby Gramacy, Mengyang Gu, and C. A. Eiffer for valuable discussions and, in some cases, feedback on earlier versions of this manuscript.

\bibliographystyle{abbrvnat}
\bibliography{references}

\end{document}